\documentclass{aa}

\usepackage{url}
\usepackage[breaklinks=true]{hyperref}
\usepackage{twoopt}
\usepackage[english]{babel}          
\usepackage{pdflscape}

\usepackage{natbib}
\bibpunct{(}{)}{;}{a}{}{,} 

\defcitealias{Morabito2014}{M14}
\defcitealias{Oonk2017}{O17}
\defcitealias{Salas2017}{S17}
\defcitealias{Salas2018}{S18}

\usepackage[utf8]{inputenc}
\usepackage[normalem]{ulem}
\usepackage{siunitx}
\usepackage{amsmath, amssymb}
\usepackage[farskip=0pt]{subfig}

\usepackage{multirow,bigstrut,ctable}
\usepackage{rotating}
\usepackage[varg]{txfonts} 
\usepackage{threeparttable}

\hypersetup{
    colorlinks,
    linkcolor={blue!50!black},
    citecolor={blue!50!black},
}

\newcommand\T{\rule{0pt}{2.6ex}}       
\newcommand\B{\rule[-1.2ex]{0pt}{0pt}} 

\newcommand{\kms}{km s$^{-1}$}

\newcommand{\asec}{\text{$^{\prime\prime}$}}

\newcommand{\calib}{\text{3C 196}}

\newcommand{\trg}{\text{3C 190}}

\newcommand{\losoto}{\texttt{LoSoTo}}
\newcommand{\dppp}{\texttt{DPPP}}
\newcommand{\wsclean}{\texttt{WSCLEAN}}
\newcommand{\aoflagger}{\texttt{AOFlagger}}

\begin{document}

\title{Searching for the largest bound atoms in space}

\author{K. L. Emig\inst{1}, P. Salas\inst{1}, F. de Gasperin\inst{1,2}, J. B. R. Oonk\inst{1,3},  M. C. Toribio\inst{4}, A. P. Mechev\inst{1}, \\ H. J. A. R\"ottgering\inst{1}, and A. G. G. M. Tielens\inst{1}}

\authorrunning{Emig et al.}

\institute{Leiden Observatory, Leiden University, P.O.Box 9513, NL-2300 RA, Leiden, The Netherlands, \email{emig@strw.leidenuniv.nl}
\and Hamburger Sternwarte, Universit\"at Hamburg, Gojenbergsweg 112, D-21029, Hamburg, Germany
\and ASTRON - the Netherlands Institute for Radio Astronomy, P.O.Box 2, NL-7990 AA, Dwingeloo, the Netherlands
\and Department of Space, Earth and Environment, Chalmers University of Technology Onsala Space Observatory, SE-439 92 Onsala, Sweden
}

\date{Received ... / Accepted ...}

\abstract
{Radio recombination lines (RRLs) at frequencies $\nu < 250$ MHz trace the cold, diffuse phase of the interstellar medium. Yet, RRLs have been largely unexplored outside of our Galaxy. Next generation low frequency interferometers, such as LOFAR, MWA and the future SKA, with unprecedented sensitivity, resolution, and large fractional bandwidths, are enabling the exploration of the extragalactic RRL universe.}
{We describe methods used to (1) process LOFAR high band antenna (HBA) observations for RRL analysis, and (2) search spectra for the presence of RRLs blindly in redshift space. }
{We observed the radio quasar 3C 190 ($z \approx 1.2$) with the LOFAR HBA. In reducing this data for spectroscopic analysis, we have placed special emphasis on bandpass calibration. We devised cross-correlation techniques that utilize the unique frequency spacing between RRLs to significantly identify the presence of RRLs in a low frequency spectrum. We demonstrate the utility of this method by applying it to existing low-frequency spectra of Cassiopeia A and M 82, and to the new observations of 3C 190.}
{RRLs have been detected in the foreground of 3C 190 at $z=1.12355$ (assuming a carbon origin), owing to the first detection of RRLs outside of the local universe (first reported in Emig et al. 2019). Towards the Galactic supernova remnant Cassiopeia  A, we uncover three new detections: (1) stimulated C$\epsilon$-transitions ($\Delta \mathsf{n} = 5$) for the first time at low radio frequencies, (2) H$\alpha$ transitions at 64 MHz with a FWHM of 3.1 \kms\, the most narrow and one of the lowest frequency detections of hydrogen to date, and (3) C$\alpha$ at $v_{\rm LSR} \approx 0$ \kms\ in the frequency range 55--78 MHz for the first time. Additionally we recover C$\alpha$, C$\beta$, C$\gamma$, and C$\delta$ from the -47 \kms\ and -38 \kms\ components. In the nearby starburst galaxy, M 82, we do not find a significant feature. With previously used techniques, we reproduce the previously reported line properties.}
{RRLs have been blindly searched and successfully identified in Galactic (to high order transitions) and extragalactic (to high redshift) observations with our spectral searching method. Our current searches for RRLs in LOFAR observations are limited to narrow ($<100$ \kms) features, owing to the relatively small number of channels available for continuum estimation. Future strategies making use of a wider band (covering multiple LOFAR subbands) or designs with larger contiguous frequency chunks would aid calibration to deeper sensitivities and broader features.}


\maketitle

\section{Introduction} 
\label{sec:intro}

Recombination lines that are observable at low radio frequencies ($\lesssim 1$ GHz) involve transitions with principal quantum numbers $\mathsf{n} \gtrsim 200$. They trace diffuse gas ($n_e \approx 0.01$ -- 1 cm$^{-3}$) that can be considered cool in temperature ($T_e \approx 10 - 10^4$ K). 

Observations in our Galaxy suggest that the most prominent radio recombination lines (RRLs) at $\nu \lesssim 250$ MHz arise from cold ($T_e \approx 10 - 100$ K), diffuse ($n_e \approx 0.01$ -- 0.1 cm$^{-3}$) gas within diffuse HI clouds and in clouds surrounding CO-traced molecular gas \citep{Roshi2011, Salas2018}. This reservoir of cold gas, referred to as ``CO-dark'' or ``dark-neutral'' gas, is missed by CO and HI emission observations. Yet it is estimated to have a comparable mass to the former two tracers \citep{Grenier2005} and is the very site where the formation/destruction of molecular hydrogen transpires. 

In addition, RRLs are compelling tools to study the physics of the interstellar medium (ISM) because they can be used to determine physical properties of the gas, specifically the temperature, density, path-length and radiation field \citep{Shaver1975a, Salgado2017b}. Pinning down these properties is key to describing the physical state of a galaxy and understanding the processes of stellar feedback. RRL modeling depends, not on chemical-dependent or star-formation modeling, but on (redshift-independent) atomic physics. Since low-frequency RRLs are stimulated transitions, they can be observed to high redshift against bright continuum sources \citep{Shaver1978b}. With evidence of stimulated emission being dominant in local extragalactic sources at $\sim$1 GHz \citep{Shaver1978a}, it is plausible that RRLs can be observed out to $z\sim4$. Low-frequency RRLs, therefore, have a unique potential to probe the ISM in extragalactic sources out to high redshift. 

The physical properties of gas can be determined when RRLs are observed over a range of principal quantum numbers. However, they are extremely faint, with fractional absorption of $\sim$10$^{-3}$ or less. At frequencies of $\sim$150 MHz, RRLs have a $\sim$1 MHz spacing in frequency. By $\sim$50 MHz, their spacing is $\sim$0.3 MHz. Large fractional band-widths enable many lines to be observed simultaneously. On one hand this helps to constrain gas properties  \citep[e.g. ][hereafter \citetalias{Oonk2017}]{Oonk2017}. On the other, it enables deeper searches through line-stacking \citep[e.g. ][]{Balser2006}.

The technical requirements needed for stimulated RRL observations can be summarized as follows: (1) large fractional bandwidths that span frequency ranges 10 -- 500 MHz for cold, carbon gas and 100 -- 800 MHz for (partially) ionized, hydrogen gas; (2) spectral resolutions of $\sim$0.1 kHz for Galactic observations and $\sim$1 kHz for extragalactic; (3) high sensitivity per channel; and (4) spatial resolutions which ideally resolve the $\lesssim$1--100 pc emitting regions. These requirements have inhibited wide-spread, in-depth studies of low-frequency RRLs in the past, largely due to the low spatial resolutions and the narrow bandwidths of traditional low-frequency instruments -- owing to the difficulty of calibrating low frequency observations affected by the ionosphere. 

However, with next generation low frequency interferometers, such as the Low Frequency Array \citep[LOFAR;][]{vanHaarlem2013}, the Murchison Widefield Array \citep[MWA;][]{Tingay2013}, and the future Square Kilometer Array (SKA), new possibilities are abound for the exploration of the ISM through RRLs. LOFAR has currently been leading the way, thanks to its raw sensitivity and the flexibility of offering high spectral and spatial resolutions. 

LOFAR operates between 10 MHz -- 90 MHz via  low band antennas (LBA) and 110 MHz -- 250 MHz via high band antennas (HBA). The array consists of simple, inexpensive dipole antennas grouped into stations. LOFAR is an extremely flexible telescope, offering multiple observing modes (beam-formed, interferometric) and  vast ranges of spectral, timing and spatial resolutions. It is the first telescope of its kind in the Northern Hemisphere and will uniquely remain so for the foreseeable future.

The first Galactic RRL analyses with LOFAR have been directed towards Cassiopeia A (Cas A), a bright supernova remnant whose line of sight intersects gas within the Perseus Arm of the Galaxy. These studies have highlighted the capability of RRL observations, and through updated modeling of atomic physics \citep{Salgado2017a,Salgado2017b}, have laid important ground work for the field in a prototypical source. It was shown that with recombination lines spanning principal quantum numbers of ${\sf n} = 257-584$ the electron temperature, density, and path-length of cold, diffuse gas can be determined to within 15 percent \citepalias{Oonk2017}. Wide bandwidth observations, especially at the lowest observable frequencies (11 MHz), can be used to constrain gas physical properties together with the $\alpha$, $\beta$ and $\gamma$ transitions of carbon in a single observation \citep[][hereafter \citetalias{Salas2017}]{Salas2017}. Through pc-resolution and comparisons with other cold gas tracers, it was shown that low-frequency RRLs indeed trace CO-dark molecular gas on the surfaces of molecular clouds \citep[][hereafter \citetalias{Salas2018}]{Salas2018}. Finally, observations towards Cygnus A demonstrated that bright extragalactic sources can also be used to conduct Galactic pinhole studies \citep{Oonk2014}.

The first extragalactic observations with LOFAR are showing that low frequency RRLs provide means to trace cold, diffuse gas in other galaxies and out to high redshift. These studies focused on M 82 \citep[][hereafter \citetalias{Morabito2014}]{Morabito2014}, a nearby prototypical starburst galaxy, and the powerful radio quasar 3C 190 at z$\sim$1.2 \citep{Emig2019a}. While these searches are important first steps that show RRL detections are possible, they also indicate that detailed analyses of stacking are necessary \citep{Emig2019a}.

In this article, we cover that much needed detailed look. We describe the methods behind the detection of RRLs in 3C 190 \citep{Emig2019a}. We explain processing of LOFAR 110--165 MHz observations for spectroscopic analysis (Section \ref{sec:dr}). We then focus on the methods used to search across redshift space for the presence of RRLs in a low-frequency spectrum (Section \ref{sec:zsearch}). We apply this technique to existing  55--78 MHz LOFAR observations of Cas A (Section \ref{sec:casa}) and demonstrate that it can be used to recover known RRLs in the spectrum, in addition to previously unknown features. We then focus on the LOFAR observations of M 82 in Section \ref{sec:m82}. 
Section \ref{sec:3c190} covers the application of our spectral search to 3C 190. We discuss the utility of the method and implications for future observations in Section \ref{sec:discuss}. Conclusions are given in Section \ref{sec:conclude}.

\begin{figure}
\includegraphics[width=0.48\textwidth]{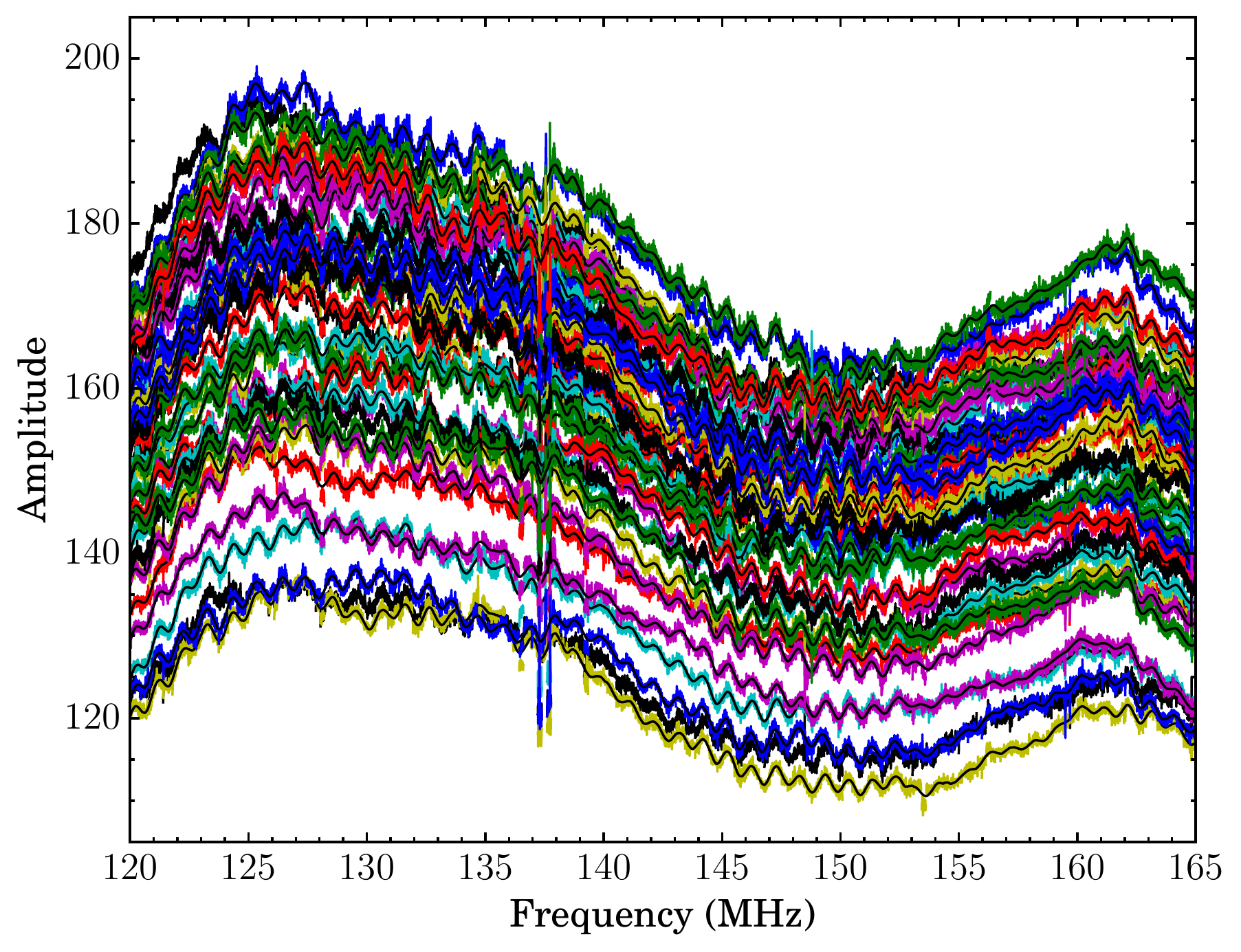}
\caption{Bandpass solutions of the XX polarization towards the primary calibrator 3C196, in which stations are represented with different colors. These solutions demonstrate the global shape of the bandpass, as well a $\sim$1 MHz ripple which results from a standing wave. Unflagged RFI is still present between 137--138 MHz. All core-stations have the same cable length and thus standing wave of the same periodicity. The fit to the solutions of each station, which is transferred to the target, is shown in black.}
\label{fig:bp}
\end{figure}

\section{Spectroscopic Data Reduction} 
\label{sec:dr}

\begin{figure*}
\includegraphics[width=\textwidth]{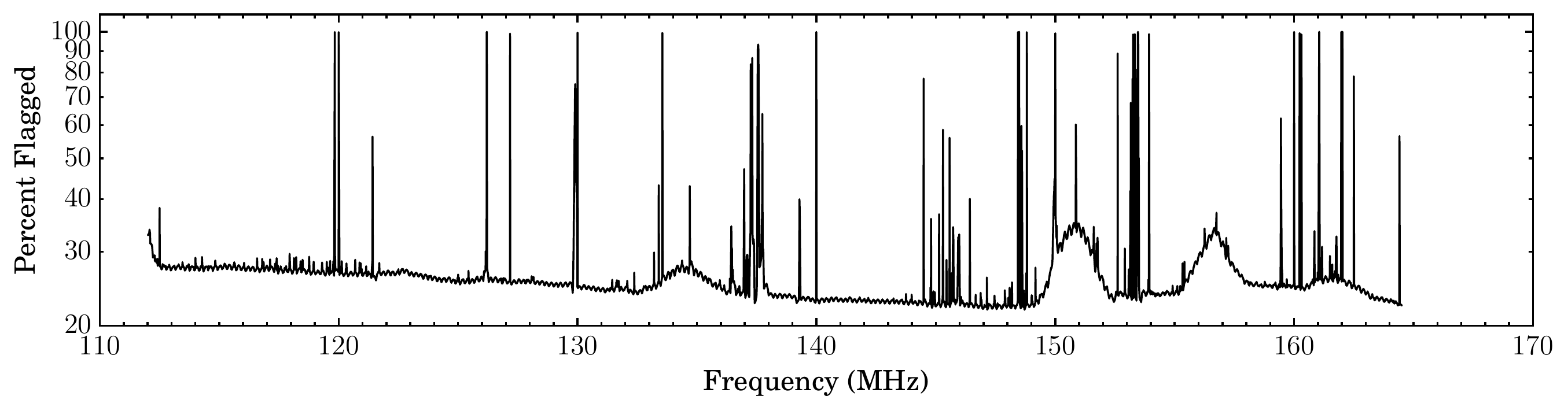}
\caption{Percentage of flagged visibilities per channel in the calibrated target data. The total percentage here includes 10 remote stations and 4 core stations that have been flagged. Broad bumps centered at frequencies of about 135, 151, 157, and 162 MHz show broad-band RFI that results from intermodulation products of DAB amplifiers.}
\label{fig:flag}
\end{figure*}

In this section, we cover the implementation of direction-independent spectroscopic calibration for LOFAR HBA(-low; 110--190 MHz) interferometric observations. Special attention is placed on the bandpass as it is one of the most crucial steps and underlies the main motivations for our strategy.

\subsection{HBA Bandpass}
\label{sec:bp}

The observed bandpass of LOFAR's HBA-low (110--190 MHz filter) is a result of the system response across frequency. It is dependent upon both hardware and software related effects. The physical structure and orientation of the dual-polarization dipole antennas influence the global shape of the bandpass. Once the sky signal enters through the antennas, an analog beamformer forms a single station beam. This beam (and the model of it) is frequency dependent (except at the phase center). The signal is transferred from the station through coaxial cables to a processing cabinet. As a result of an impedance mismatch between the cables and receiver units, standing waves are imprinted in the signal, in proportion to the length of the cable. Standing waves causing the $\sim$1 MHz ripple are apparent in Figure \ref{fig:bp}. An analog filter is then applied; this is responsible for the roll-off at either end of the bandpass. Next, after analog-to-digital conversion, a polyphase filter (PPF) is applied to split the data into subbands 195.3 kHz wide, which each have fixed central frequencies (for a given filter and digital-converter sampling frequency).
The data are transported to the off-site correlator via optical fibers. A second PPF is applied, now to each subband, to re-sample the data into channels. This PPF imprints sinusoidal ripples within a subband. This can and is corrected for by the observatory. However, after the switch to the COBALT correlator in 2014, residuals of the PPF are present at the $\sim$1\% level. This PPF is also responsible for in-subband bandpass roll-off which renders edge channels unusable. In-subband roll-off together with fixed subband central frequencies results in spectral observations that are processed with non-contiguous frequency coverage.

Radio frequency interference (RFI) is another major contributor to frequency dependent sensitivity. Of particular hindrance that has increased over the past years \citep[for comparison, see][]{Offringa2013} is digital audio broadcasting (DAB). DABs are broadcast in 1.792 MHz wide channels in the frequency ranges 174-- 195 MHz. However, with the output power reaching non-linear proportions ($\propto P^{3}$), intermodulation products of the DABs appear at frequencies of 135, 151, 157, and 162 MHz, as can be seen in Figure \ref{fig:flag}. The sensitivity reached and the stability of the bandpass is affected in these regions, rendering data unusable for RRL studies at the central most frequencies.

\subsection{Procedure}
\label{sec:dr_proc}

In this section we describe the reduction of 3C 190 observations. An overview of the steps we take in our data reduction strategy include: flagging and RFI removal of calibrator data, gain and bandpass solutions towards the primary calibrator, flagging and RFI removal of target data, transfer of bandpass solutions to the target, self-calibrated phase and amplitude corrections towards the target, and  imaging. These steps are described in detail below.

Processing was performed with the SURFSara Grid processing facilities\footnote{https://www.surf.nl} making use of LOFAR Grid Reduction Tools \citep{Mechev2017, Mechev2018}. It relies on the LOFAR software \dppp\ \citep{VanDiepen2018}, \losoto\ \citep{DeGasperin2019}, \wsclean\ \citep{Offringa2014}, and \aoflagger\ \citep{Offringa2012} to implement the necessary functions.

The observations of \trg\ were taken with the LOFAR HBA-low on Jan 14, 2017 (Project ID: \texttt{LC7\_027}). The set up was as follows: four hours were spent on \trg, with ten minutes on the primary calibrator \calib~before and after. The 34 stations of the Dutch array took part in the observations. Frequencies between 109.77 MHz and 189.84 MHz were recorded and divided into subbands of 195.3125 kHz. Each subband was further split into 64 channels and recorded at a frequency resolution of 3.0517 kHz. While taken at 1 s time intervals, RFI removal and averaging to 2 s were performed by the observatory before storing the data.

\subsubsection{Pre-processing and flagging}
\label{sec:preproc}

Before calibration, we first implemented a number of flagging steps. Using \dppp, we flagged the calibrator measurement sets for the remote station baselines, keeping only the 24 core stations (CS). Since similar ionospheric conditions are found above stations this close in proximity \citep{Intema2009, deGasperin2018}, this heavily reduces direction dependent effects, and this also avoids the added complication of solving for the sub-microsecond drifting time-stamp of the remote stations \citep[e.g.][]{VanWeeren2016}.

As the CS of the HBA are split into two ears, we filtered out the ear-to-ear cross-correlations. Four channels (at 64 channel resolution) at both edges of each subband are flagged to remove bandpass roll-off. With \aoflagger, we used an HBA-specific flagging strategy to further remove  RFI. We flagged all data in the frequency ranges 170 MHz -- 190 MHz due to the DABs (see Section \ref{sec:bp}). Additionally we flagged stations CS006HBA0, CS006HBA1, CS401HBA0, CS501HBA1 due to bandpass discontinuity. The data were then averaged in time and in frequency to  6 s and 32 channels per subband (or 6.1034 kHz).

\subsubsection{Calibration solutions towards the primary calibrator}
\label{sec:cal}

With the scan of \calib\ (ObsID: L565337) taken at the start of the observation, we first smoothed the visibilities and weights with a Gaussian weighting scheme that is proportional with one over the square of the baseline length \citep[e.g. see ][]{DeGasperin2019}. With \dppp, we obtained diagonal (XX and YY) gain solutions towards \calib, at full time resolution and with a frequency resolution of 1 subband, while filtering out baselines shorter than 500 m to avoid large-scale sky emission. An 8 component sky-model of 3C 196 was used (courtesy of A. Offringa).

Next, we collected the solution tables from each subband and imported them into \losoto. For each channel, we found its median solution across time, the results of which are shown in Figure \ref{fig:bp}. After 5$\sigma$ clipping, we fit the amplitude vs. frequency solutions with a rolling window (10 subbands wide) polynomial (6th order). With a window of 10 subbands, we attempted to fit over subband normalization issues, interpolate over channels which were flagged or contained unflagged RFI --- e.g. RFI-contaminated channels between 137--138 MHz in Figure~\ref{fig:bp} --- and avoid transferring per channel scatter to the target. The fit to these solutions is also shown in Figure~\ref{fig:bp}.

\subsubsection{Calibration and imaging of the target Field}
\label{sec:trg}

Flagging and averaging of the target data were performed as described in Section \ref{sec:preproc}. We then applied the bandpass solutions found with the primary calibrator \calib. We next smoothed the visibilities with the baseline-dependent smoother. Considering that ionospheric effects were minimal, the CS are all time-stamped by the same clock, and our target \trg\ is a bright and dominant source, we solved explicitly for the diagonal phases with \dppp\ with a frequency resolution of one subband and at full time resolution, while filtering out baselines shorter than 500 m. We performed this self-calibration using a Global Sky Model \citep{vanHaarlem2013}, which included 128 sources down to 0.1 Jy within a 5 degree radius. We then applied these solutions to full resolution data (32 channels, 6 s).

To correct for beam errors and amplitude scintillation, we performed a ``slow'' amplitude correction. We first averaged the data down to a 30 s time-resolution, then smoothed the visibilities with the baseline-based weighting scheme. While again filtering out baselines shorter than 500 m, we used \dppp\ to solve the amplitude only, every 30 sec and twice per subband. Before applying this amplitude correction, the solution tables from each subband are imported into \losoto. Using \losoto, we clipped outliers and smoothed the solutions in frequency-space with a Gaussian of full width half maximum (FWHM) covering four subbands. These solutions are then applied to full resolution data (32 channels per subband, 6 s).

Our last step was to create an image for each channel. With \wsclean, a multi-frequency synthesis image was created for each subband, from which the clean components are extracted and used to create channel images of greater depth. Channel images were created with Briggs 0.0 weighting out to 11x11 square degrees field of view.  We convolved every channel image to the same resolution of 236\asec, a few percent larger than the lowest-resolution image, using \texttt{CASA} \citep{McMullin2007}. The flux density was then extracted from a fixed circular aperture of diameter 236\asec, and a spectrum was created for each subband.

\section{Searching RRLs in redshift space} 
\label{sec:zsearch}

The second main focus of this paper covers our method to search for RRLs blindly across redshift space. RRLs may not be detected individually, but wide-bandwidth observations enable detections as a result of stacking. Since the frequency spacing between each recombination line is unique ($ \nu \propto \mathsf{n}^{-2}$), a unique redshift can be blindly determined with the detection of two or more lines. In stacking across redshift space, there are two main issues that require caution. The first is the low $N$ statistics involved in the number of lines (10 -- 30 spectral lines in HBA, 20--50 in LBA) used to determine the stack averaged profile. The second is the relatively small number of channels available to estimate the continuum in standard (64 channels or less per subband) LOFAR observations. The method we employ does not depend on the unique set up of LOFAR and can be applied to observations with other telescopes. 

The main steps of the method include:
\begin{itemize}
\item[1.] assume a redshift and stack the spectra at the location of availables RRLs; repeat for a range of redshifts (see Section~\ref{sec:stackproc})
\item[2.] cross-correlate the observed spectrum in optical depth units with a template spectrum populated with Gaussian profiles at the location of the spectral lines for a given redshift; repeat over a range of redshifts (see Section~\ref{sec:spec_cc})
\item[3.] cross-correlate the integrated optical depth of the stacked spectrum across redshift with the integrated optical depth of a template spectrum across redshift in order to corroborate mirror signals (see Section~\ref{sec:stack_cc})
\end{itemize}

We compare the values of the normalized cross-correlations, and identify outliers assuming a normal distribution. Here we note that a single cross-correlation value is not necessarily meaningful in itself, but it is the relative comparison of the cross-correlation values across redshifts which identifies outliers.  We require both cross-correlations result in a 5$\sigma$ outlier at each redshift, as deemed necessary from simulated spectra (Section~\ref{sec:synthspec}). Once a significant feature is identified by these means, we subtract the best fit of the RRL stack. We then repeat the procedure to search for additional transitions or components. In the sections below, we describe each step in further detail.

\subsection{Stacking RRLs}
\label{sec:stackproc}

We began processing the spectra by flagging. We manually flagged subbands with clearly poor bandpasses. We Doppler corrected the observed frequencies as Doppler tracking is not supported by LOFAR. We flagged additional edge channels that are affected by bandpass roll-off. Before removing the continuum, we flagged and interpolated over channels for which $>$50\% of the visibility data were flagged as well as channels with a flux density greater than five times the standard deviation.

For a given redshift, we blanked the channels (assuming a certain line-width) at the expected frequency of the line when estimating the continuum. For stimulated transitions at low frequencies, we have that $I_{\mathrm{line}} \approx I_{\mathrm{cont}} \tau_{\mathrm{line}}$, where the intensity extracted from the observations is $I_{\rm obs} \approx I_{\mathrm{line}} + I_{\mathrm{cont}}$. Therefore, subtracting a continuum fit and dividing by it resulted in the optical depth, $(I_{\rm obs} - I_{\rm cont}) / I_{\rm cont} = \tau_{\rm line}$. The continuum was fit with a 1st or 2nd order polynomial, chosen based on the $\chi^2$ of the fit. Considering $\chi^2$ of the fit and the rms of the subband, we flagged subbands which have outlying values. 

Taking the central frequency of each subband, we found the RRL closest in frequency and used it to convert frequency units into velocity units. We then linearly interpolated the velocities to a fixed velocity grid, which has a frequency resolution equal to or greater than the coarsest resolution of all subbands. We weight subbands by their rms ($w = \sigma^{-2}$). We then stack-averaged all of the $N$ subbands available, where the optical depth of each channel is given by 
\begin{equation}
<\tau_{\mathrm{chan}}> = \frac{ \Sigma_{i=0}^N (w_i \tau_i) }{ \Sigma_{i=0}^N w_i  }
\label{eq:weighted_tau}
\end{equation}
and $i$ represents each subband going into the stack. An effective frequency, $\nu_{\rm eff}$, of the stacked spectral line was determined from the weighted mean frequencies among the lines stacked. We determined an effective principal quantum number, ${\sf n}_{\rm eff}$, by taking the integer quantum number of the line that was closest in frequency to $\nu_{\rm eff}$.

The error of each channel of the stacked spectra reflects the standard deviation of a weighted mean, which is $\sigma_{<\rm chan >} = (\Sigma_{i=0}^N w_i )^{-1/2}$. When comparing the integrated optical depth at each redshift, we integrated within a region one half of the blanked region and centered at $v_{\rm stack} = 0$ \kms. The spectral noise per channel of the stack at each redshift is determined by taking the weighted standard deviation of all channels outside the line-blanking region. 

\begin{figure*}
    \centering
    \includegraphics[width=0.75\textwidth]{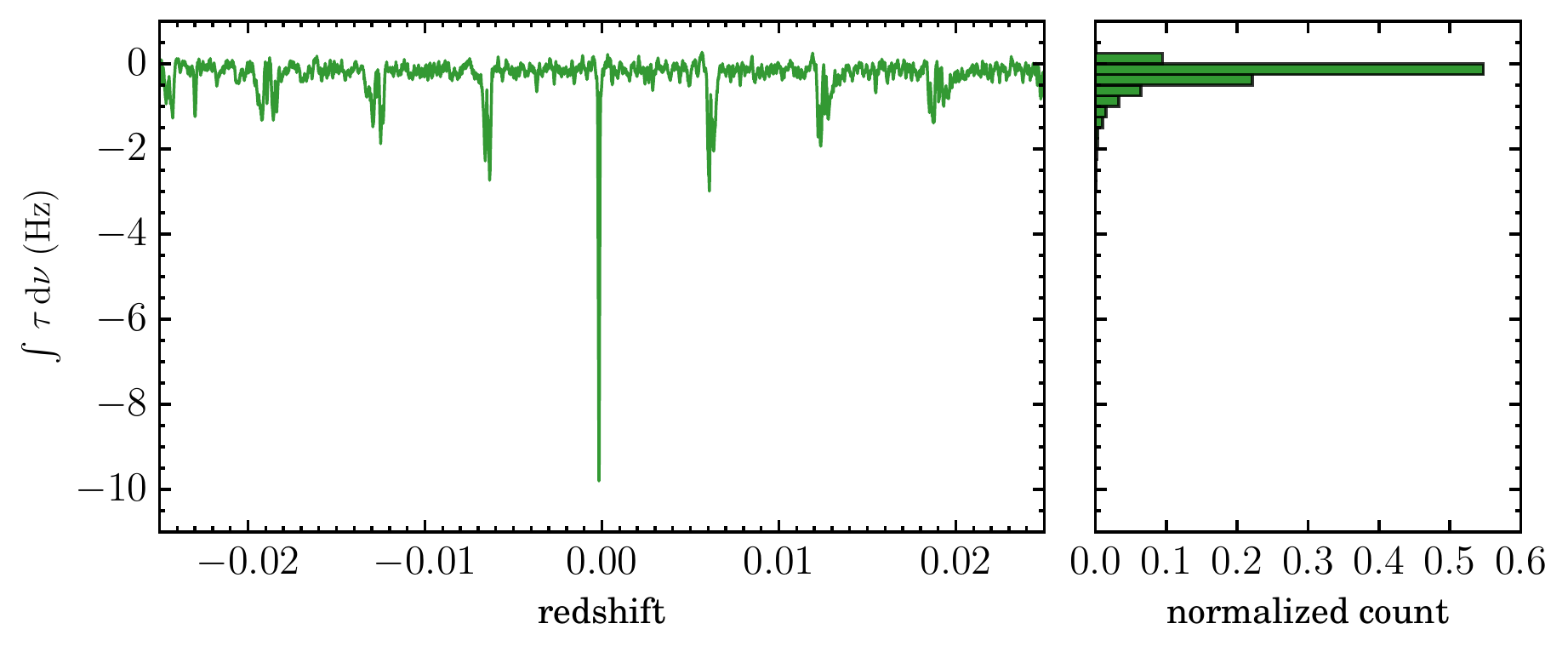}
    \caption{Cas A spectra has been stacked for C$\alpha$ RRLs across a range of redshifts. The plot on the {\it left} shows the optical depth integrated at the central velocities (-7.7 \kms\ to +19.8 \kms, which encompasses the -47 and -38 \kms\ components) at each redshift. Here we clearly see an outlying peak at the redshift ($z=-0.000158$) of the -47 \kms\ component. The histogram on the {\it right} shows the binned distribution of integrated optical depth. Cas A $\alpha$-transition stacking demonstrates our method in a high signal-to-noise regime.}
    \label{fig:alpha_itau}
\end{figure*}

We show an example of stacking across redshift in Figure \ref{fig:alpha_itau}.

\subsection{Spectral cross-correlation}
\label{sec:spec_cc}

\begin{figure*}
    \centering
    \includegraphics[width=0.75\textwidth]{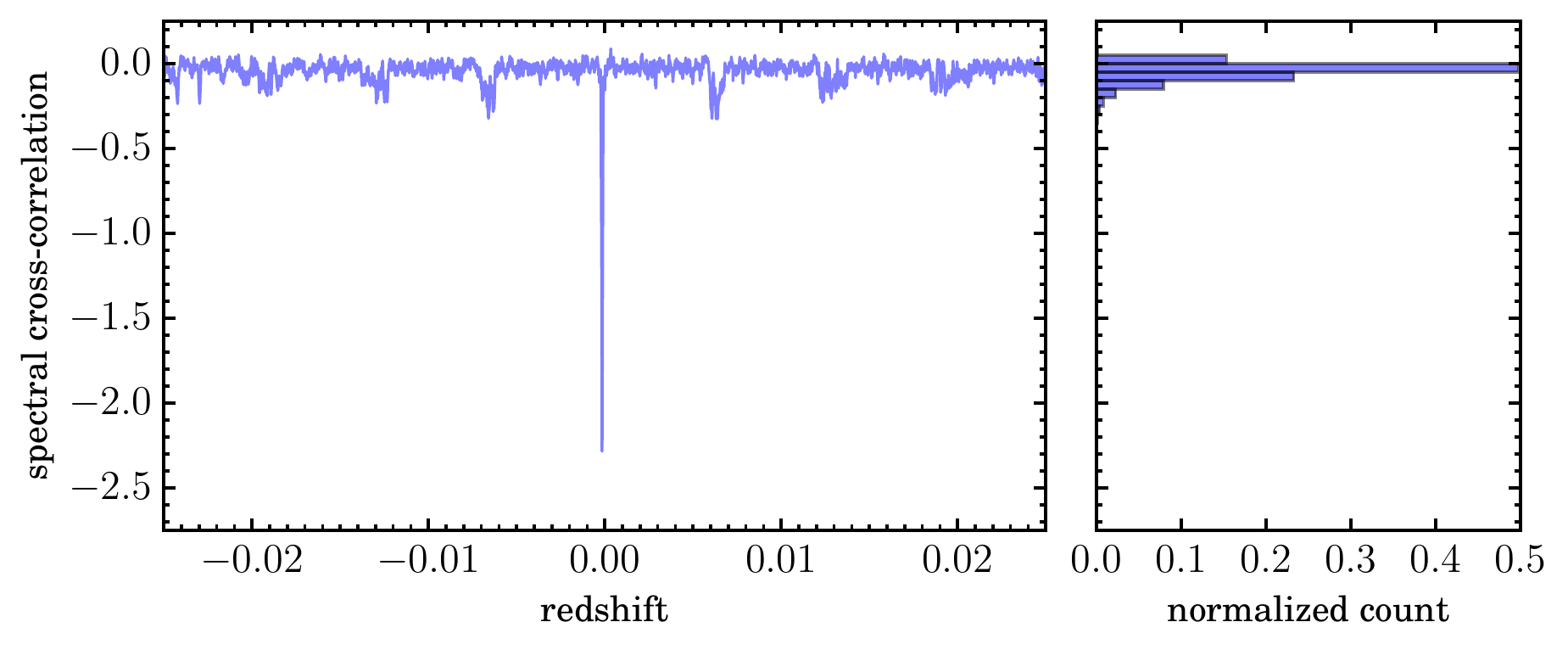}
    \caption{Cas A spectra have been cross-correlated for C$\alpha$ RRLs with a template spectrum across a range of redshifts. The plot on the {\it left} shows the spectral cross-correlation value at each redshift. Again we clearly see an outlying peak at the redshift ($z=-0.000158$) of the -47 \kms\ component. The histogram on the {\it right} shows the binned distribution of cross-correlation values. Cas A $\alpha$-transition stacking demonstrates our method in a high signal-to-noise regime.}
    \label{fig:alpha_ccspec}
\end{figure*}

For each redshift, we performed a cross-correlation between a template spectrum and the observed spectrum, both in units of optical depth and frequency. The template spectrum was populated with Gaussian line profiles at the location of the spectral lines that contributed to the final stack. The line profiles were normalized to a peak optical depth of unity and their FWHM was set by half the width of the line-blanked region. We then took the cross-correlation and normalized it proportionally with the number of lines that went into the stack, i.e. the total area under the template spectrum. This was the procedure that \cite{Morabito2014} implemented for a single redshift, except we included a normalization since the number of lines stacked at each redshift did not remain constant. We show an example of cross-correlating the spectrum across a range of redshifts in Figure \ref{fig:alpha_ccspec}.

\subsection{Stack cross-correlation}
\label{sec:stack_cc}

With a second cross-correlation, we corroborate ``mirrors'' of the signal that can be found at $(\Delta z)_{\rm mirror} =  \overline{ \Delta \nu_{\mathsf{n,n+N}} } / \nu_{\mathsf{n}}$, or multiples thereof, where $\nu_{\mathsf{n}}$ is the frequency of the effective $\mathsf{n}$-level of the stack, and $\overline{ \Delta \nu_{\mathsf{n,n+N}}}$ is the average of the change in frequency between the effective $\mathsf{n}$ and all of the $N$ other $\mathsf{n}$-levels. Since the difference in frequency spacing between $\alpha$-transitions $\mathsf{n}$ and $\mathsf{n} + 1 $ is small ($\sim$1 \%), mirrors of the feature, which are more broadly distributed (in velocity space) and reduced in peak intensity, occur at offsets in redshift that match the frequency spacing between adjacent lines.

\begin{figure*}
    \centering
    \includegraphics[width=\textwidth]{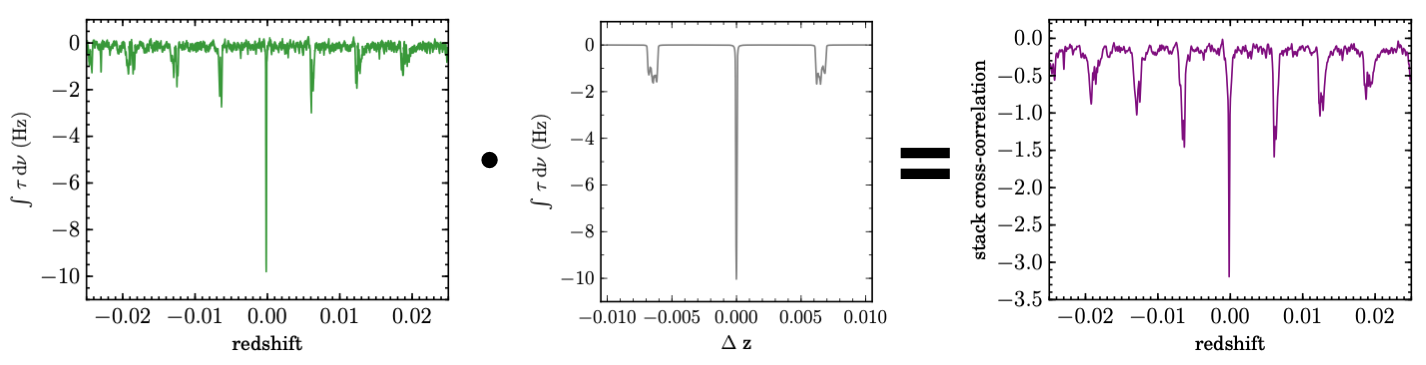}
    \caption{Here we demonstrate the stack cross-correlation. Cas A spectra have been stacked for C$\alpha$ RRLs across a range of redshifts. The plot on the {\it left} shows the optical depth integrated at the central velocities at each redshift. The {\it middle} plot shows the integrated optical depth resulting from stacking the template spectrum. It has been done for a redshift of $z=0.000158$. We cross-correlate the left plot with the middle plot to obtain the stack cross-correlation on the {\it right}. }
    \label{fig:ccdemo}
\end{figure*}

\begin{figure*}
    \centering
    \includegraphics[width=0.75\textwidth]{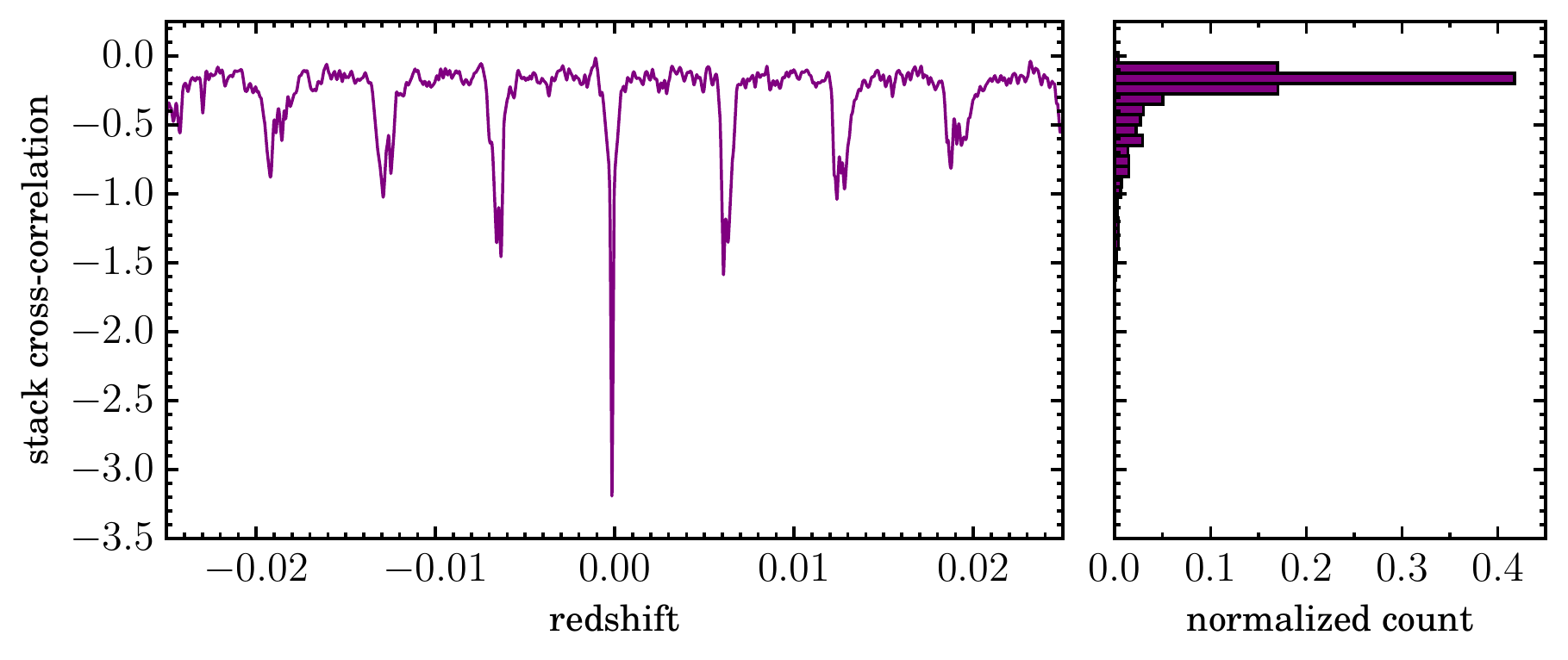}
    \caption{The integrated optical depth of the observed Cas A spectra is cross-correlated with the expected template spectrum. Here we clearly see an outlying peak at the redshift ($z=-0.000158$) of the -47 \kms\ component. The histogram on the {\it right} shows the binned distribution of integrated optical depth. Cas A $\alpha$-transition stacking demonstrates our method in a high signal-to-noise regime.}
    \label{fig:alpha_ccstack}
\end{figure*}

We identified redshifts with an outlying ($>4\sigma$) value in the normalized spectral cross-correlation. For those redshifts, we took its template spectrum (see Section \ref{sec:spec_cc}), stacked the template lines at an assumed redshift, $z_{\rm cen}$, and integrated the optical depth. We then stacked and integrated the template spectrum at a range of redshifts covering $\approx z_{\rm cen} \pm (\Delta z)_{\rm mirror}$. We then cross-correlated (a) the integrated optical depth of the template stack as a function of redshift, with (b) the observed integrated optical depth of the data at each redshift (see Figures \ref{fig:ccdemo} \& \ref{fig:alpha_ccstack}). We found it best to restrict $(\Delta z)_{\rm mirror}$ such that only one mirror of the feature was present. 

\begin{figure}
    \centering
    \includegraphics[width=0.47\textwidth]{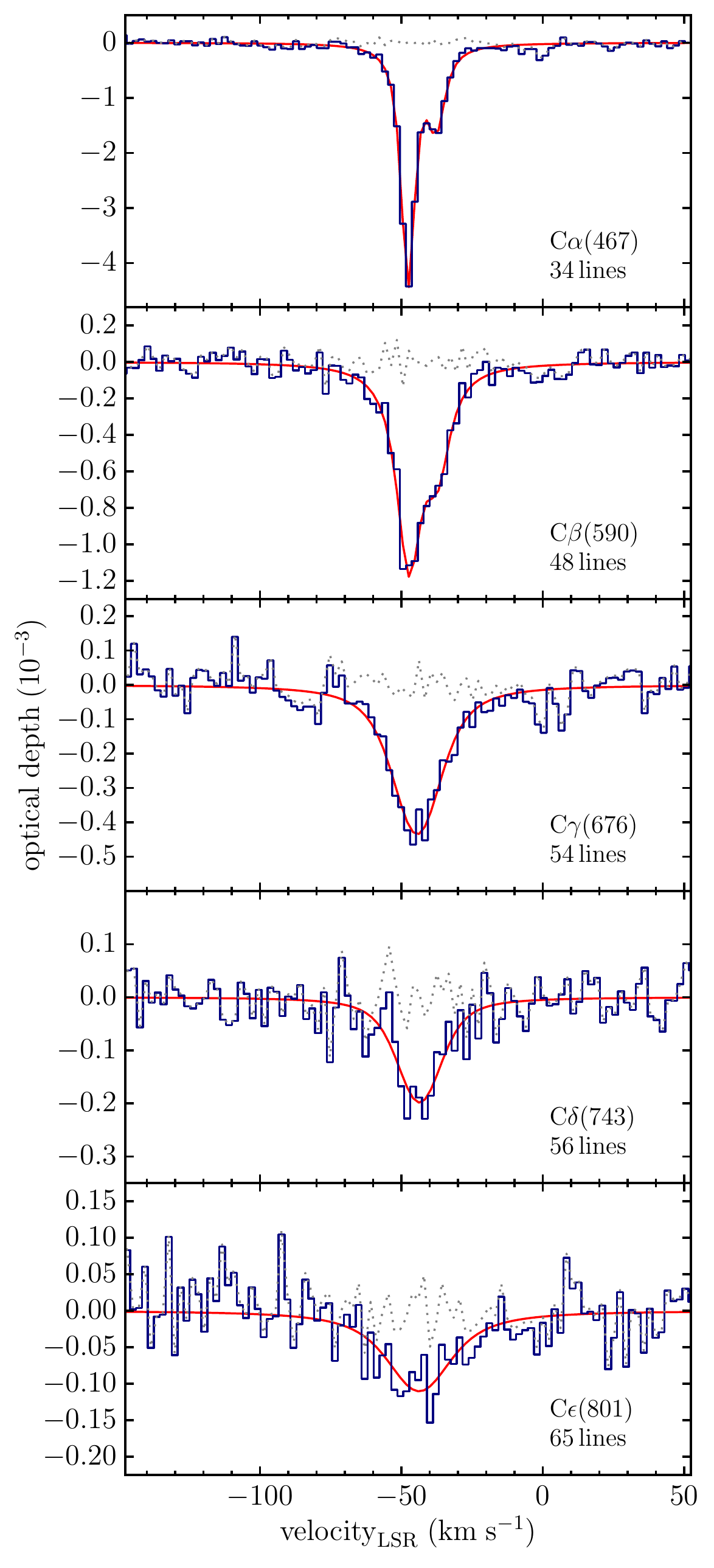}
    \caption{The stack averaged profiles for the carbon transitions associated with the $v_{\rm LSR} = -47$ \kms\ and $v_{\rm LSR} = -38$ \kms\ components in 55 MHz -- 78 MHz observations of Cas A. These have all been significantly identified by our method.}
    \label{fig:casa_bestfits}
\end{figure}

\begin{figure}
    \centering
    \includegraphics[width=0.48\textwidth]{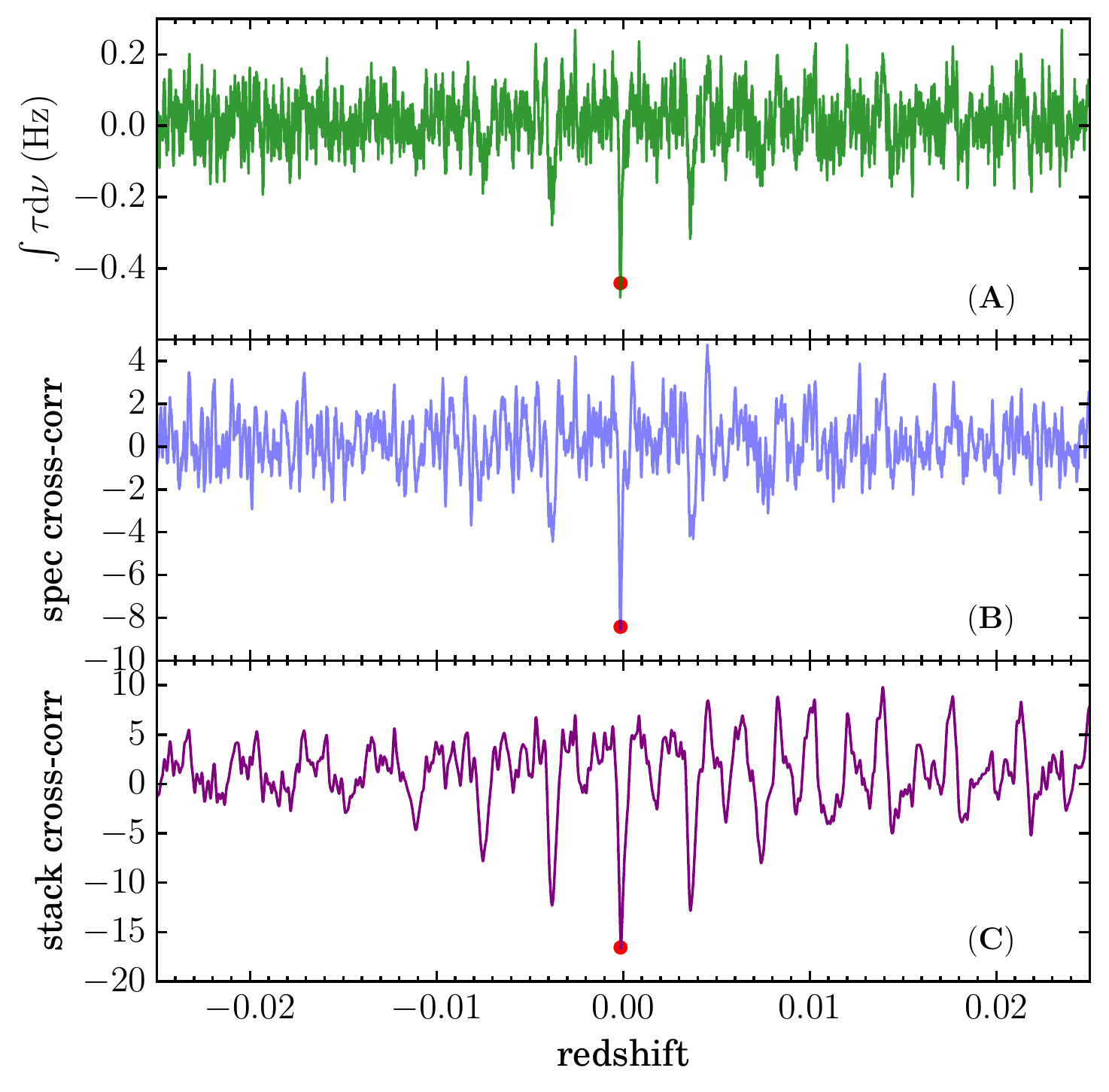}
    \caption{The three steps of our method applied to $\epsilon$-transition stacking in the spectrum of Cas A: {\bf (A)} the integrated optical depth at each redshift (see Section \ref{sec:stackproc}) where ``mirrors'' are seen at e.g. $z_m = \pm 0.004$, {\bf (B)} the spectral cross-correlation (see Section \ref{sec:spec_cc}), and {\bf (C)} the stack cross-correlation (see Section \ref{sec:stack_cc}). This example demonstrates the behavior of the method in the presence of broad RRLs. }
    \label{fig:casa_epsilon}
\end{figure}

\subsection{Validation with Synthetic Spectra}
\label{sec:synthspec}

We developed this procedure first on synthetic spectra. The synthetic spectra were constructed by injecting Gaussian noise, distributed about an optical depth of 0. Radio recombination line profiles were populated at frequencies covering 110 -- 160 MHz. Fifteen trials were made; the trials covered a variety of physical conditions (line properties), noise (signal-to-noise) regimes, and redshifts out to $z=2$. Note, we did not insert continuum flux and started the procedure from the continuum subtraction stage. The tests were focused on a low signal-to-noise regime where continuum estimation and removal do not substantially affect the noise properties of the stacked line results.

Results with these synthetic spectra showed that radio recombination lines which resulted in a stacked line profile whose Gaussian fit had a peak signal-to-noise of 2.7$\sigma$ could be successfully recovered in the cross-correlations at the 5$\sigma$ level. The results of these tests also showed that it was essential for both the spectral and the stack cross-correlations to achieve a 5$\sigma$ outlier at the input redshift. When these criteria were satisfied no false positives were recovered.

\section{Cassiopeia A}
\label{sec:casa}

\begin{table*}
    \centering
    \begin{tabular}{cccccccc}
    \hline
    Line &$\mathsf{n}$ range  &Frequency  &Line center  &Lorentz FWHM  &Total FWHM  &$\int\tau\,\mathrm{d}\nu$ \T \\ 
        &       &(MHz)      &(\kms)     &(\kms)     &(\kms)     &(Hz) \B \\
    \hline
    C$\alpha$(467)  &436 -- 489  &64.38  &$-47.6\pm0.9$  &$5.0\pm0.3$  &$6.1\pm0.7$  &$6.7\pm0.8$ \T \\
                    &            &       &$-37.7\pm1.0$  &$6.7\pm1.1$  &$7.8\pm1.7$  &$3.7\pm0.9$ \\
    C$\alpha$(469)    &436 -- 489  &63.41  &$-0.5\pm1.2$  & -  &$3.6\pm0.5$  &$0.23\pm0.05$  \\                
    C$\beta$(590)  &549 -- 616  &63.79  &$-47.3\pm1.1$  &$10.0\pm0.5$  &$10.7\pm0.8$  &$5.8\pm0.5$ \\
                   &            &       &$-37.2\pm1.3$  &$9.9\pm1.1$  &$10.7\pm1.5$  &$2.7\pm0.5$ \\
    C$\gamma$(676)  &628 -- 705  &63.53  &$-44.4\pm1.2$  &$14.2\pm2.3$  &$22.8\pm3.1$  &$7.8\pm1.1$ \\
    C$\delta$(743)  &691 -- 775  &63.77  &$-43.6\pm1.5$  &$11.4\pm1.7$  &$21.2\pm3.2$  &$2.8\pm0.5$ \\
    C$\epsilon$(801)  &744 -- 835  &63.53  &$-43.7\pm2.0$  &$22.2\pm3.2$  &$29.4\pm3.0$  &$3.7\pm0.5$ \B \\
    \hline
    \end{tabular}
    \caption{Properties of the carbon line profiles that have been detected in the spectrum of Cas A. The uncertainties quoted are 1$\sigma$.}
    \label{tab:casa}
\end{table*}

The line-of-sight towards the supernova remnant Cas A has been the focus of detailed investigations of low-frequency RRLs in our Galaxy. We make use of LOFAR spectra between 55 MHz -- 80 MHz first presented in \citetalias{Oonk2017}. The flux densities ($\sim 2\times 10^4$ Jy at 55 MHz) were extracted from a 14 x 14 arcmin$^2$ aperture, covering the entire source. The spectra have a channel width of 0.381 kHz (512 channels per subband), which corresponds to velocity resolutions between 1.4 \kms\ -- 2.1 \kms\ over this frequency range. Starting from spectra in units of flux density that had been Doppler corrected to the Local Standard of Rest (LSR) reference frame, we implemented flagging as described in Section \ref{sec:stackproc}. Since the velocities of the components are known before hand, we directly specified the line-blanking regions then estimated the continuum using the line-free regions. With spectra in optical depth units, we iterated through the steps of our method. This differs from the implementation of M82 and 3C 190 in that it does not include continuum subtraction for each redshift. During stacking, we interpolate to a spectrum with a velocity resolution of 2.1 \kms, and we cover the redshift ranges $-0.025 \leq z \leq 0.025$, Nyquist sampling redshift intervals, $\delta z_{\rm sample} = 3 \times 10^{-6}$, which corresponds to about 0.9 \kms.

We initiated the procedure by stacking for the most prominent spectral lines in the spectrum, the C$\alpha$ transitions of the -47 \kms\ and -38 \kms velocity components associated with gas in the Perseus Arm of our Galaxy. Two components are clearly distinguishable, but they overlap with one another. Therefore we defined a line-blanking region (-7.7 \kms, 19.8 \kms) that was centered on the most prominent component, -47 \kms, but encapsulated both components.  The integrated optical depth which resulted from stacking for C$\alpha$ RRLs at each redshift is shown in Figure \ref{fig:alpha_itau}. The redshift ($z_{-47 \rm km/s} = -0.000151$) corresponding to $v_{\rm LSR}$ = -47 \kms\ is most prominent. Mirrors of the signal are clearly visible at multiples of $z_m \approx \pm 0.007$; they degrade in peak intensity the larger the multiple of $z_m$. Figure \ref{fig:alpha_ccspec} shows the spectral cross-correlation. The template spectrum has been populated with two Voigt fits (see Figure \ref{fig:casa_bestfits}, Table \ref{tab:casa}) to the -47 \kms\ and -38 \kms\ components rather than Gaussians with a FWHM in proportion to the line-blanked region, as line-broadening due to radiation and pressure broadening (for more details see Sec. \ref{sec:casa_c}) are present in these lines \citepalias{Oonk2017}.

We stack the template spectrum over a redshift range of $z_{-47 kms} \pm$ 0.01, as shown in Figure \ref{fig:ccdemo}. The result of cross-correlating this template function with the integrated signal vs redshift  of the true spectrum is shown in Figure \ref{fig:alpha_ccstack}. In this high signal-to-noise test case, all three steps of the method identify the -47 \kms\ component with large significance. To subtract this component (and make more sensitive searches for additional components), we grouped the spectral lines into six different stacks and subtracted the best fit at the location of the spectral line. Grouping lines into six stacks minimized residuals due to slow changes in line properties at different frequencies. 
We then repeated the procedure, stacking for C$\beta$, C$\gamma$, C$\delta$,  C$\epsilon$,  C$\zeta$, and again C$\alpha$. After each transition was tested for and significantly identified, we subtracted the best fit line profile, and then continued with testing the next transition.

\subsection{Carbon RRL Results}
\label{sec:casa_c}

We significantly identify the presence of six unique transitions associated with the -47 (and -38) \kms\ component(s) of the Perseus Arm of the Galaxy and one transition associated with a 0 \kms velocity component in the Orion Arm of the Galaxy. In addition to C$\alpha$ associated with the -47 and -38 \kms\ components, we show the detections of C$\beta$, C$\gamma$, and C$\delta$ at this frequency for the first time. As a first for low-frequency RRLs, we have also significantly identified ($7.8\sigma$) the C$\epsilon$ transition. The stacked spectra and line profiles are shown in Figure \ref{fig:casa_bestfits}. We show the steps of our method that recover the $\epsilon$-transitions in Figure \ref{fig:casa_epsilon}. The properties of all Cas A line profiles can be found in Table \ref{tab:casa}. The errors of each quantity were determined from the variance of each variable as determined by the fit. When fitting the $\beta$ lines, the Gaussian width was fixed to the values derived from the $\alpha$ lines: 1.2 \kms\ and 1.1 \kms\ for the -47 and -38 \kms\ components, respectively. Additionally, the best fit of the $\gamma$ line was used to set the Gaussian width of the $\delta$ and $\epsilon$ lines at 6.0 \kms since the two velocity components are blended for these transitions.

The CRRLs towards Cas A have been exhaustively analyzed in \citetalias{Oonk2017}, \citetalias{Salas2017}, and \citetalias{Salas2018}. From the data available at $\mathsf{n} < 580$ for the -47 \kms\ gas component, spanning roughly 30 MHz -- 1 GHz, densities and temperatures have been determined to be $n_e(-47) = 0.04 \pm 0.005$ cm$^{-3}$, $T_e(-47) = 85 \pm 5$ K with a path-length of $L_{C+}(-47) = 35.3 \pm 1.2$ pc \citepalias{Oonk2017}. Similarly the -38 \kms\ component was found to have the properties $n_e(-38) = 0.04 \pm 0.005$ cm$^{-3}$, $T_e(-38) = 85 \pm 10$ K, and $L_{C+}(-38) = 18.6 \pm 1.6$ pc. For principal quantum numbers $\sf{n}>550$ line blending has not allowed for the two velocity components to be analyzed independently.  \citetalias{Salas2017} analyzed CRRLs present at 10--33 MHz from the blend of these velocity components. Slightly less dense gas with a somewhat stronger radiation field dominated the absorption lines at very low frequencies: $T_e = 60-98$ K, $T_{r,100} = 1500-1650$ K and $n_e = 0.02-0.035$ cm$^{-3}$. \citetalias{Salas2017} note the difference between the RRLs observed at low and high quantum number could arise from (1) a variation in the spectral index of the background continuum source across the 5$'$ face, weighting the gas differently at higher and lower frequencies, (2) different cloud depths and thus the single slab with which the gas was model would not hold, and (3) the physical conditions of the -38 and -47 \kms\ components may not be the same. 

\begin{figure}
    \includegraphics[width=0.48\textwidth]{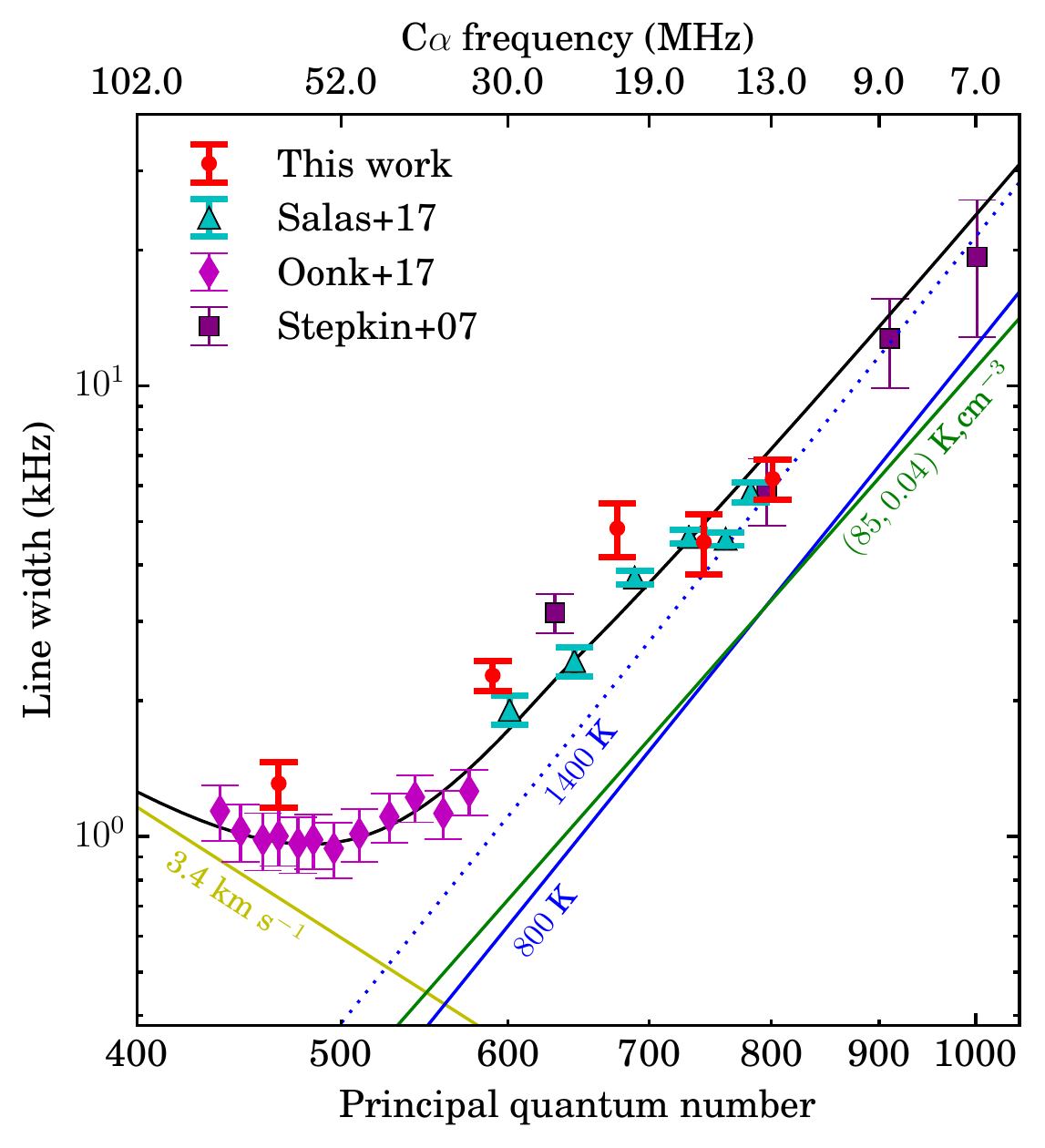}
    \caption{Line-width for the -47 \kms\ velocity component (or the sum of blended -47 and -38 \kms\ components) as a function of principal quantum number. The red points show the measured line-widths from this work (Table \ref{tab:casa}); the C$\alpha(467)$ and C$\beta(590)$ are derived only from the -47 \kms\ velocity component, while the higher order transitions represent a single fit to the blended -47 and -38 \kms\ components. The cyan triangles show the line-widths of the C$\alpha$ and C$\beta$ transitions of the -47 \kms\ velocity component from \citetalias{Salas2017}. The purple diamonds show the line-widths of the -47 \kms\ component from \citetalias{Oonk2017}. The purple squares show the C$\alpha$, C$\beta$, C$\gamma$ and C$\delta$ data, for which the -47 and -38 \kms\ components are blended,  from \cite{Stepkin2007}. The colored lines show the contribution from Doppler broadening (yellow line), pressure broadening (green line) and radiation broadening (blue lines). The solid black line shows the model which best fit the line-widths from \citetalias{Salas2017}: a Doppler line-width of 3.4 \kms, $T_e = 85$ K, $n_e = 0.04$ cm$^{-3}$ and a radiation field which is a combination of a power law with $T_{r,100} = 800$ K and $\alpha = -2.6$ plus a contribution to the radiation field from Cas A.}
    \label{fig:line-width}
\end{figure}

Here, we focus on the new data uncovered by our stacking procedure. In Figure \ref{fig:line-width}, we compare the observed line-width of the CRRLs as a function of principal quantum number with models for line broadening due to the Doppler effect, pressure broadening, as well as radiation broadening \citepalias{Salas2017}. The line-widths of $\alpha$ and $\beta$ lines are measured from only the -47 \kms\ component, where as the line-widths of the $\gamma$, $\delta$, and $\epsilon$ lines are a blend of the -47 and -38 \kms\ components. Our results show that the $\alpha$, $\beta$, and $\gamma$ line-widths are slightly over-estimated compared with  prior measurements, although well within error. The broad line widths we measure are likely contaminated by higher order transitions which have not been subtracted from the data and by the blending of the two velocity components. In the more detailed analyses of \citetalias{Salas2017} (also applied to \citetalias{Oonk2017}) and \cite{Stepkin2007}, the spectral lines were subtracted and an additional baseline correction was performed before re-stacking for the final spectra. \citetalias{Salas2017} goes on to validate the line-width measurement error and the integrated optical depth error by applying the approach to synthetic spectra. They confirmed that for principal quantum numbers ${\sf n} < 800$ the line properties were reproduced accurately to within 16 percent; this also applies to the \citetalias{Oonk2017} results. Since a validation of this kind goes beyond the scope of this paper, we underline the greater certainty in the detailed analysis of the \citetalias{Salas2017} measurements.

\begin{figure}
    \includegraphics[width=0.48\textwidth]{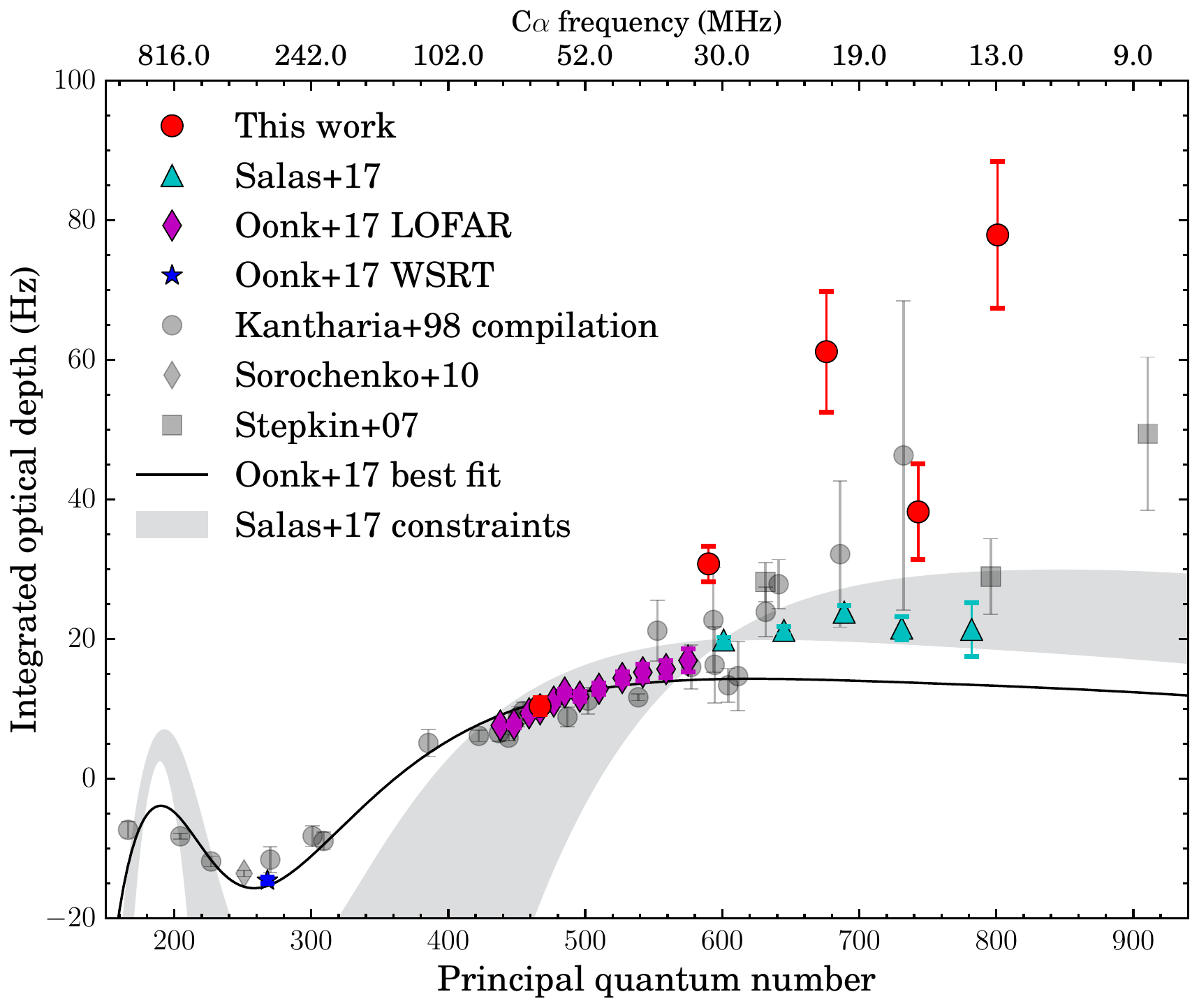}
    \caption{Integrated optical depth as a function of principal quantum number for the sum of the Perseus arm components at -47 and -38 \kms adapted from \citetalias{Salas2017}. The cyan triangles show the 10--33 MHz LOFAR observations of \citetalias{Salas2017}. The purple diamonds represent the 33--78 MHz LOFAR detections and the blue star shows the WSRT detection \citepalias{Oonk2017}. Pre-LOFAR literature data are shown in gray data points \citep{Kantharia1998, Stepkin2007, Sorochenko2010}. The black solid line is the best-fitting model from \citetalias{Oonk2017}. The gray shaded region covers the models which correspond to the physical constraints from \citetalias{Salas2017}. }
    \label{fig:itau_calpha}
\end{figure}

\begin{figure*}
    \centering
    \includegraphics[width=0.49\textwidth]{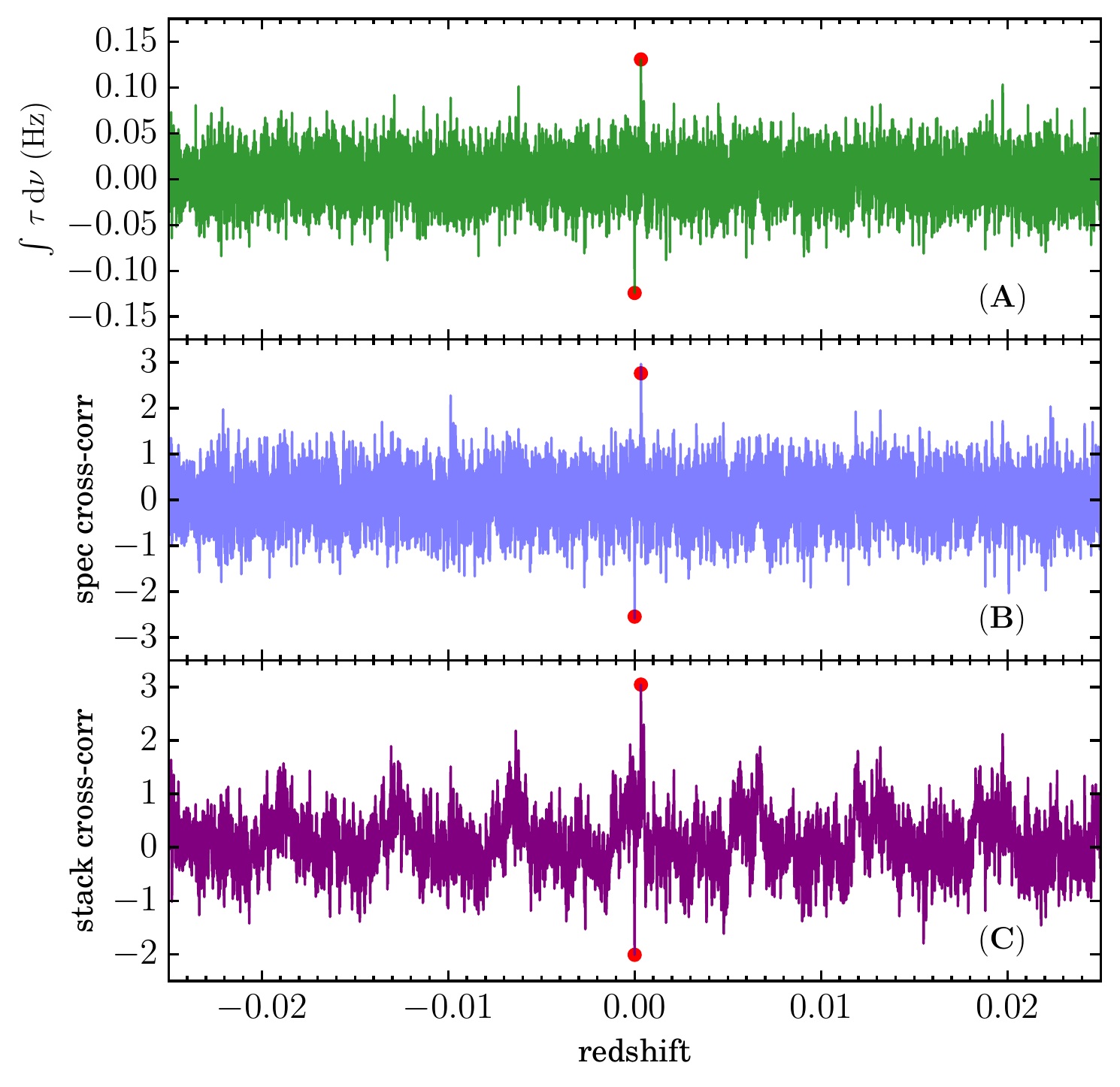}
    \includegraphics[width=0.49\textwidth]{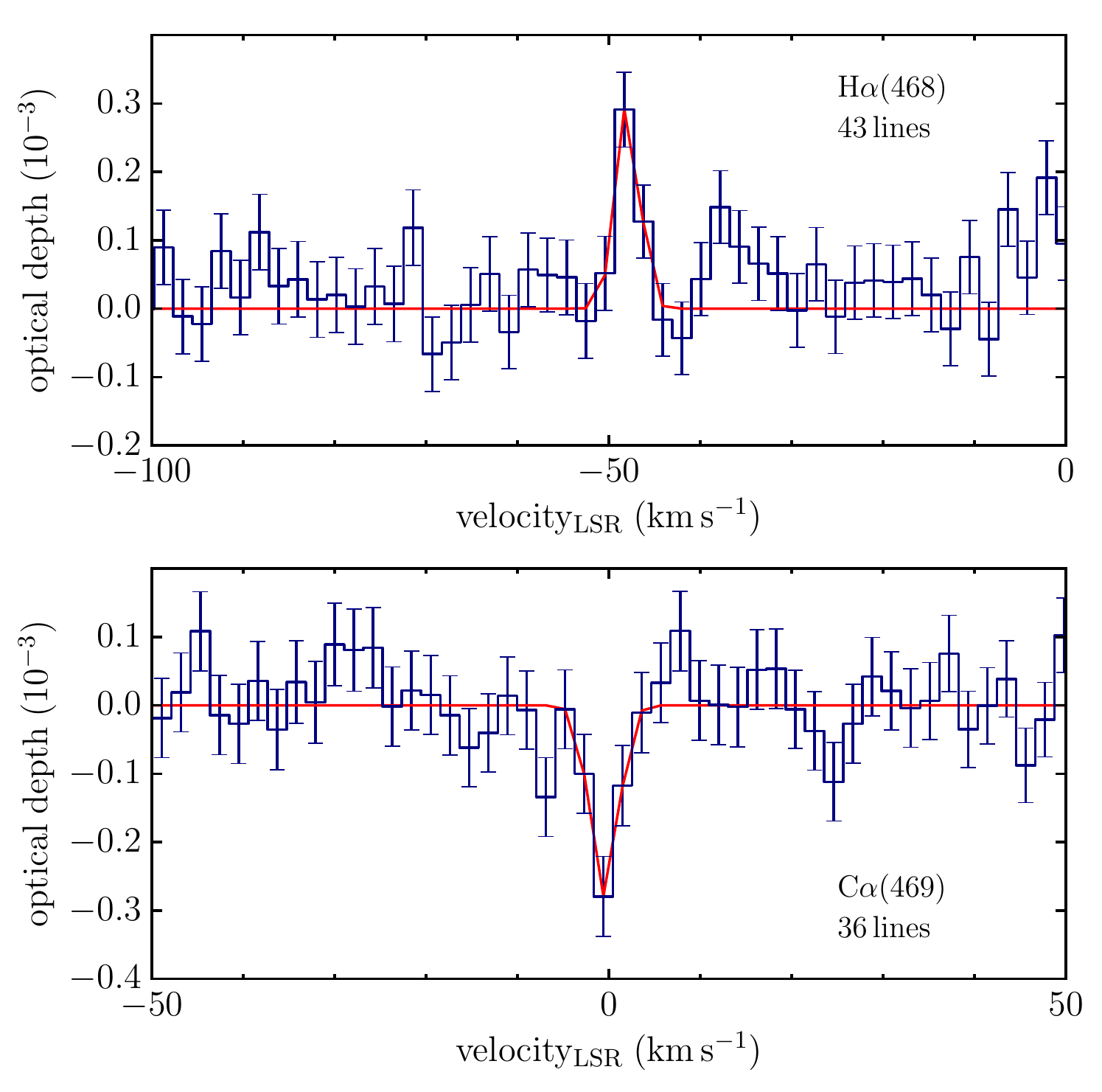}
    \caption{{\it Left:} Same as in Figure \ref{fig:casa_epsilon}, except applied to C$\alpha$-transitions in Cas A spectra. Fits to the -47 \kms\ and -38 \kms components (and all higher order transitions) have been subtracted in the spectrum of Cas A. Residuals remaining from imperfect subtraction of the brightest $\alpha$ transitions from -47 \kms\ and -38 \kms components can be see in {\bf (C)}. These results demonstrate the behavior of the method in the presence of narrow, low signal-to-noise spectral lines. The redshifts ($z_H = 0.000337$ and $z_C = -0.000002$, with respect to stacking for C$\alpha$) of the two detections are shown in red. {\it Right:} The stacked spectrum of H$\alpha$ (top) associated with the -47 \kms\ component, one of the lowest frequency detections of Hydrogen RRLs. C$\alpha$ (bottom) associated with the Orion Arm, detected at this frequency for the first time.}
    \label{fig:casa_h_c0_alpha}
\end{figure*}

Figure \ref{fig:itau_calpha} shows the integrated optical depth as a function of principal quantum number together with model constraints, adapted from the literature compilation of \citetalias{Salas2017}. The integrated optical depth reflects the sum of -47 and -38 \kms components. We converted the integrated optical depth of higher order transitions into an equivalent $\alpha$ optical depths through $I_{\Delta \mathsf{n}=1} = I_{\Delta \mathsf{n}} \frac{ \Delta \mathsf{n} \cdot M_{\Delta \mathsf{n}} }{ M_{\Delta \mathsf{n}=1} }$, where $\Delta  \mathsf{n} = \mathsf{n}_{'} - \mathsf{n}$ and $M_{\Delta \mathsf{n}}$ is the approximate oscillator strength \citep{Menzel1968}. We note that at high $\sf{n}$ involved here, difference in non-LTE effects are considered negligible ($\sim 1$\%). The integrated optical depths that we measure of the higher order ($\Delta  \mathsf{n}>1$) transitions generally lie above existing values. Again, since the maximum error induced by the processing procedure has been quantified in \citetalias{Salas2017}, we defer to those measurements. Our values are most likely over-estimated primarily due to residuals of e.g. $\alpha$ (see Figure \ref{fig:casa_h_c0_alpha}) and  $\beta$ spectral lines which were not perfectly subtracted. The scatter in the literature for ${\sf n} \gtrsim 500$ reflects the difficulty in determining the baseline of the continuum in broad, overlapping spectral lines. 

Lastly, we turn our attention to the C$\alpha$ component at $v_{\mathrm{LSR}} = 0$ \kms\ associated with the Orion Arm. While it was not reported by \citetalias{Oonk2017} at these frequencies, we significantly identify it ($5.2\sigma$) here with a rather faint optical depth of $\int \tau {\rm d}\nu = (0.23 \pm 0.05)$ Hz (see Figure \ref{fig:casa_h_c0_alpha}). Additional line properties can be found in Table \ref{tab:casa}. This result demonstrates the utility of our stacking methods for a narrow line in a low signal-to-noise regime. In Figure \ref{fig:casa_h_c0_alpha} {\bf (C)}, we see that low-level residuals from the imperfect subtraction of the bright -47 \kms\ component is present in the stack cross-correlation. 

There are now five detections in total of CRRLs associated with the 0 \kms\ component. A component at this velocity was clearly visible in WSRT P-band data at ${\sf n} = 267$ and in LOFAR 33 -- 45 MHz data (${\sf n}\sim 588$) \citepalias{Oonk2017}. \cite{Stepkin2007} also find an $\alpha$ and $\beta$ line centered at $v_{\rm LSR} = -1.6$ \kms, blended with the -47 \kms component in their low 26 MHz observations using the UTR-2 telescope.  Visual comparisons of this component with other cold diffuse gas tracers, such as $^{13}$CO(1-0) and [CI] \citepalias{Salas2018} and  HI \citep{Payne1989}, provide additional evidence for a cold neutral medium (CNM) component. We compared detections of the -47 \kms\ CRRL component with these in the Orion Arm by looking at the integrated optical depth as a function of principal quantum number. After scaling the Orion detections by a factor of 20 (to match values at ${\sf n}\sim500$), there are indications that the transition from absorption to emission occurs at higher $\sf{n}$ in this component, indicating that the gas may be less dense and/or have a warmer radiation field (e.g. see constraints of Figure \ref{fig:itau_calpha}). However, the re-scaled values and errors of these low signal-to-noise detections are within $3\sigma$ of the -47 \kms\ component and therefore consistent with its derived properties. Deep observations, particularly at higher frequencies where the line is in emission, would provide useful constraints for future modeling.

\subsection{Hydrogen RRL Results}
\label{sec:casa_h}

We also report the detection of an H$\alpha$ emission line as a result of stacking 43 lines at an average frequency of 64.08 MHz, one of the lowest frequency detections to date \citep[see also ][towards the Galactic Center at 63 MHz]{Oonk2019} and a valuable probe of the cold partially ionized phase of the ISM. This feature (see Table \ref{tab:halpha} for line properties) has been significantly identified in each step of our method (see Figure \ref{fig:casa_h_c0_alpha}), most prominently in the stack cross-correlation with a $5.4\sigma$ confidence. Figure \ref{fig:casa_h_c0_alpha} shows the feature identified at a redshift $z =0.000337$ or $v_{\rm LSR}=101$ \kms\ as these stacks are relative to C$\alpha$ RRL, for which the H$\alpha$ transitions are regularly separated by 149.4 \kms. With respect to H$\alpha$ RRLs, the central velocity is $v_{\rm LSR}= -47.9 \pm 1.1$ \kms. We find a feature of $\int \tau \mathrm{d}\nu = -0.21 \pm 0.04$ Hz, consistent with a $3\sigma$ upper limit of -0.42 Hz in this frequency range \citepalias{Oonk2017}. 

\begin{table}
    \centering
    \begin{tabular}{ccccc}
    \hline
    $\sf{n}$    &Frequency  &Line center    &FWHM   &$\int\tau\,\mathrm{d}\nu$ \T \\
                &(MHz)      &(km s$^{-1}$)  &(km s$^{-1}$)  &(Hz) \B \\
    \hline
    250     &418.35 &$-47.9\pm0.2$  &$4.6\pm0.3$    &$-1.98\pm0.24$ \T \\   
    267     &343.7  &$-47.4\pm0.14$ &$3.81\pm0.34$ &$-1.96\pm0.15$ \\   
    309     &222   &-              &-              &$-1.15\pm0.58$ \\   
    468     &64.08  &$-47.9\pm1.1$  &$3.1\pm0.5$    &$-0.21\pm0.04$ \B \\ 
    \hline
    \end{tabular}
    \caption{Hydrogen RRL detections towards Cas A. In this work we report the H$\alpha$(468) detection. We also provide the parameters of H$\alpha$(250) \citep{Sorochenko2010},  H$\alpha$(267) \citepalias{Oonk2017}, and H$\alpha$(309) \citep{Oonk2015}. }
    \label{tab:halpha}
\end{table}

The line-width is narrow (FWHM $=3.1 \pm 0.5$ \kms) in comparison to our channel resolution (2.1 \kms); considering that we have interpolated the spectra to a fixed velocity grid, it is plausible that the peak of the line is underestimated and the width is overestimated, while preserving the integrated optical depth. However, we note that our line-width is consistent within error of the FWHM of lines at higher frequencies (Table \ref{tab:halpha}). 

The measured line-widths indicate that pressure (and radiation) broadening are not dominant effects to the line-width at this frequency, since they would cause an increase in line-width towards lower frequencies. Instead the roughly constant line-width indicates a Doppler broadened profile. Assuming a purely Doppler broadened line-profile, an upper limit on the gas temperature is given by $T_e < 2 \times 10^4 {\rm K} \left( \frac{30.25 \, {\rm km \, s^{-1}}}{ \Delta v} \right)^2  $  \citep[Equation 4.7,][]{Brocklehurst1972} assuming hydrogen gas, where $\Delta v$ is the FWHM in units of \kms. We derive an upper limit of $T_e < 210 \pm 0.5$ K, directly attributing this feature to the cold neutral medium. 

We can also place a strict upper limit on the electron density if we assume the line-width is set by collisional broadening. Under this assumption, we can solve for the electron density in units of cm$^{-3}$ as $n_e = \Delta \nu_{\rm col} \left( 10^a {\sf n}^{\gamma} / \pi \right)^{-1}$, where $\Delta \nu_{\rm col}$ is the FWHM in units of Hz, $\sf n$ is the principal quantum number, and $a$ and $\gamma$ depend on the gas temperature \citep{Salgado2017b}. We refer to \cite{Salgado2017b} where the collisional coefficients were tabulated as $a = -9.620$ and $\gamma = 5.228$ for a temperature of $T_e = 200$ K, a temperature which is within 5\% of our upper limit from the Doppler profile. We find an upper limit to the electron density of $n_e < 0.10 \pm 0.02 $ cm$^{-3}$.

HRRLs at this frequency were searched for in \citetalias{Oonk2017} but went undetected. Nonetheless, with their higher frequency detection and two literature values, the HRRL gas properties were modeled. \citetalias{Oonk2017} found that the same physical conditions which best described the cold, diffuse CRRL gas ($T_e =$ 85 K and $n_e = 0.04$ cm$^{-3}$)  did not best fit the HRRL emitting gas. Alternatively, the models suggested that the hydrogen RRLs are arising from a colder and denser gas ($T_e =30-50$  K and $n_e = 0.065-0.11$ cm$^{-3}$). 

\begin{figure}
    \centering
    \includegraphics[width=0.48\textwidth]{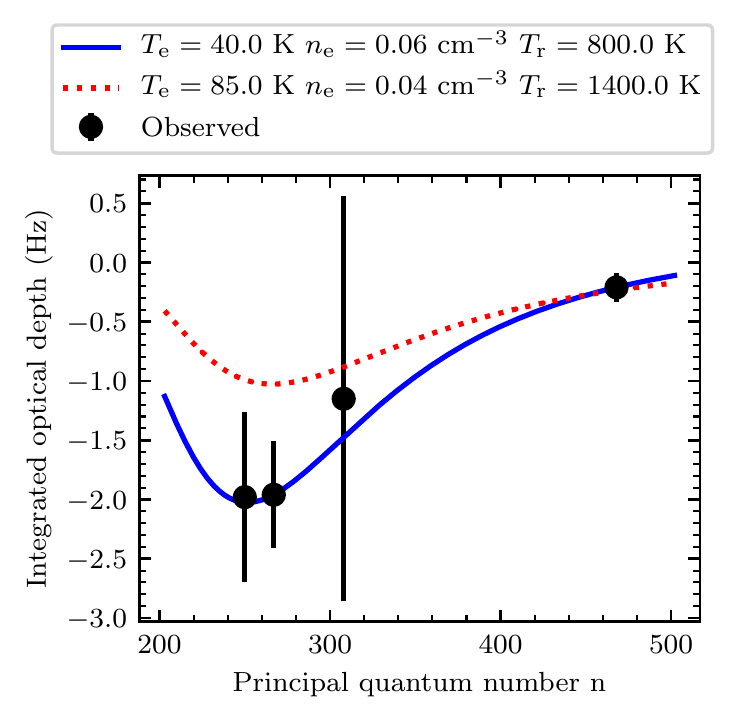}
    \caption{The integrated optical depth as a function of principal quantum number, $\sf{n}$, for the Hydrogen RRLs detections associated with the -47 \kms\ velocity component towards Cas A. The figure shows the values from Table \ref{tab:halpha} using 3$\sigma$ error bars. The blue, solid line shows the best fit model. The red, dotted line shows the best fit model \citepalias{Oonk2017} to Carbon RRL detections spanning quantum numbers 225 -- 550 (e.g. see Figure \ref{fig:itau_calpha}).}
    \label{fig:itau_hrrl}
\end{figure}

\begin{figure}
    \centering
    \includegraphics[width=0.48\textwidth]{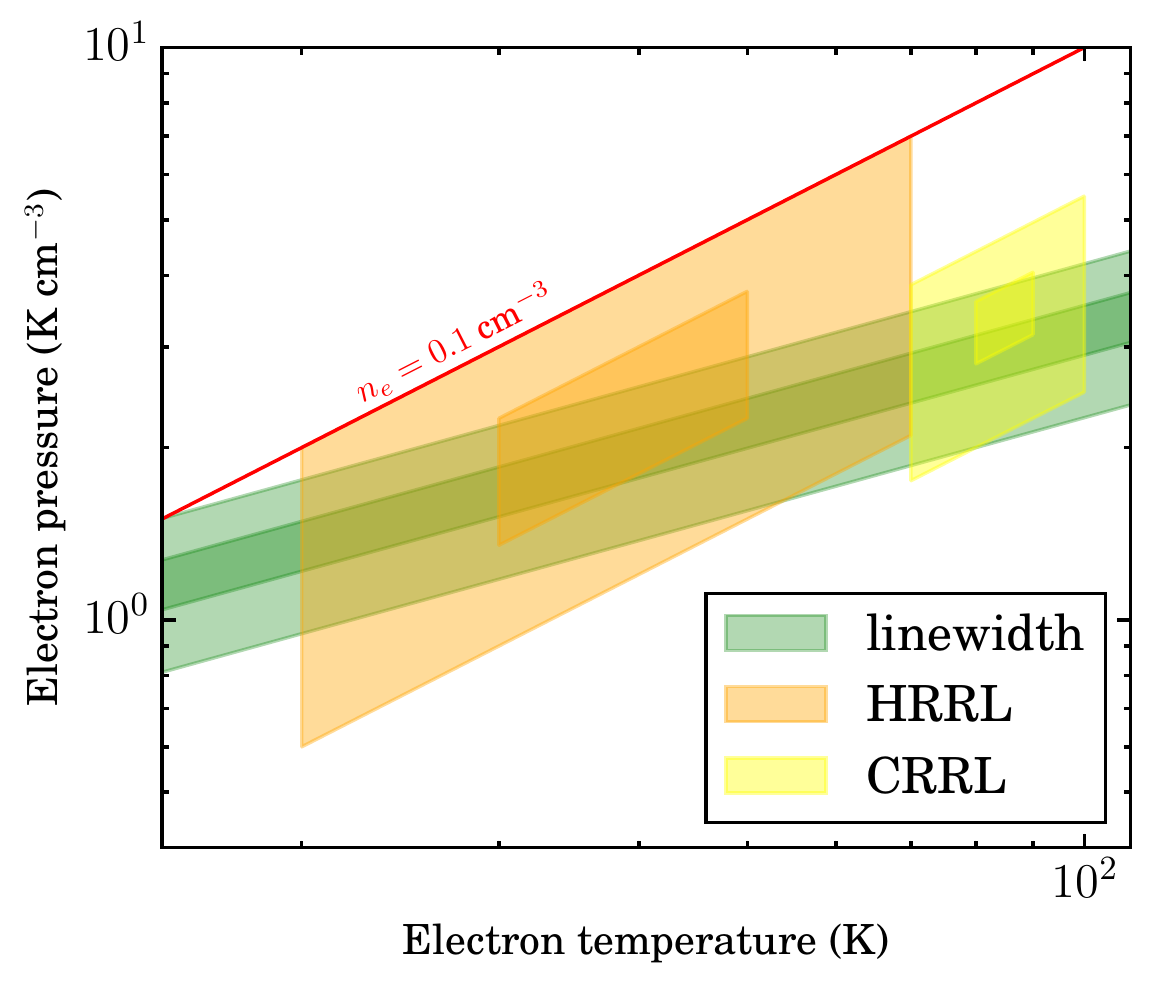}
    \caption{The electron pressure as a function of electron temperature of the -47 \kms\ velocity component towards Cas A. The red solid line shows the upper limit of a Pressure broadened line-width. Shaded regions represent 1$\sigma$ and 3$\sigma$ uncertainty. The green shaded regions show the line-width constraints from the combined Doppler, pressure and radiation broadening terms for $T_{r,100} = 1400$ K \citepalias{Salas2017}. The orange shaded regions show the physical properties of the HRRLs constrained from modeling the integrated optical depth. The yellow shaded regions show the physical properties of the CRRLs constrained from modeling the integrated optical depth \citepalias{Oonk2017}. This plot shows the distinction between the HRRLs and CRRLs, where HRRLs arise from slightly colder, denser regions in the cloud.}
    \label{fig:hrrl_pt}
\end{figure}

We re-model the physical conditions (with the procedure described in \citetalias{Oonk2017}) to derive updated constraints on the physical properties of the HRRL gas (see Figure \ref{fig:itau_hrrl}). The results provide further evidence for a component with distinctly different conditions than the CRRL gas: an electron temperature of $T_e =30-50$ K, an electron density of $n_e = 0.045-0.75$ cm$^{-3}$, an emission measure of $EM_{H+} = 0.00064 - 0.0018$ pc cm$^{-6}$, and a radiation temperature of $T_{r,100} = 800 - 2000$ K. In Figure \ref{fig:itau_hrrl}, we plot the best fit --- $T_e = 40$ K, $n_e = 0.06$ cm$^{-3}$, and $EM_{H+} = 0.0012$ pc cm$^{-6}$ --- as a function of integrated line strength and principal quantum number. Figure \ref{fig:hrrl_pt} plots all line-width and modelling constraints of the HRRL gas together with the CRRL properties.

Under the assumptions that (1) HRRLs and CRRLs originate from the same gas phase and (2) HRRLs are only ionized by cosmic rays (CR), constraints on the CR ionization rate can be determined from the ratio of the Hydrogen and Carbon RRL integrated optical depths \citep[e.g.][]{Neufeld2017, Sorochenko2010}.  The results from our models challenge assumption (1). While the line-widths and central velocities of the two tracers do not require different phases, it is possible that the gas motion is perpendicular to the line-of-sight. A high spatial resolution (2$'$) analysis of the CRRLs towards Cas A \citepalias{Salas2018} highlighted additional relevant effects: there are variations in peak line-strength by as much as a factor of seven within this region, and through comparisons with other cold gas tracers, a spatially resolved transition of the gas from a diffuse atomic state to a dense molecular cloud was identified. Spatially resolving HRRLs across the region would allow us to structurally locate the emission and how its intensity is distributed. In conclusion, a higher resolution analysis of HRRLs is needed to determine the validity of assumptions (1) and (2) in constraining the CR ionization rate.

\section{M 82} 
\label{sec:m82}

\begin{figure*}
    \centering
    \includegraphics[width=0.48\textwidth]{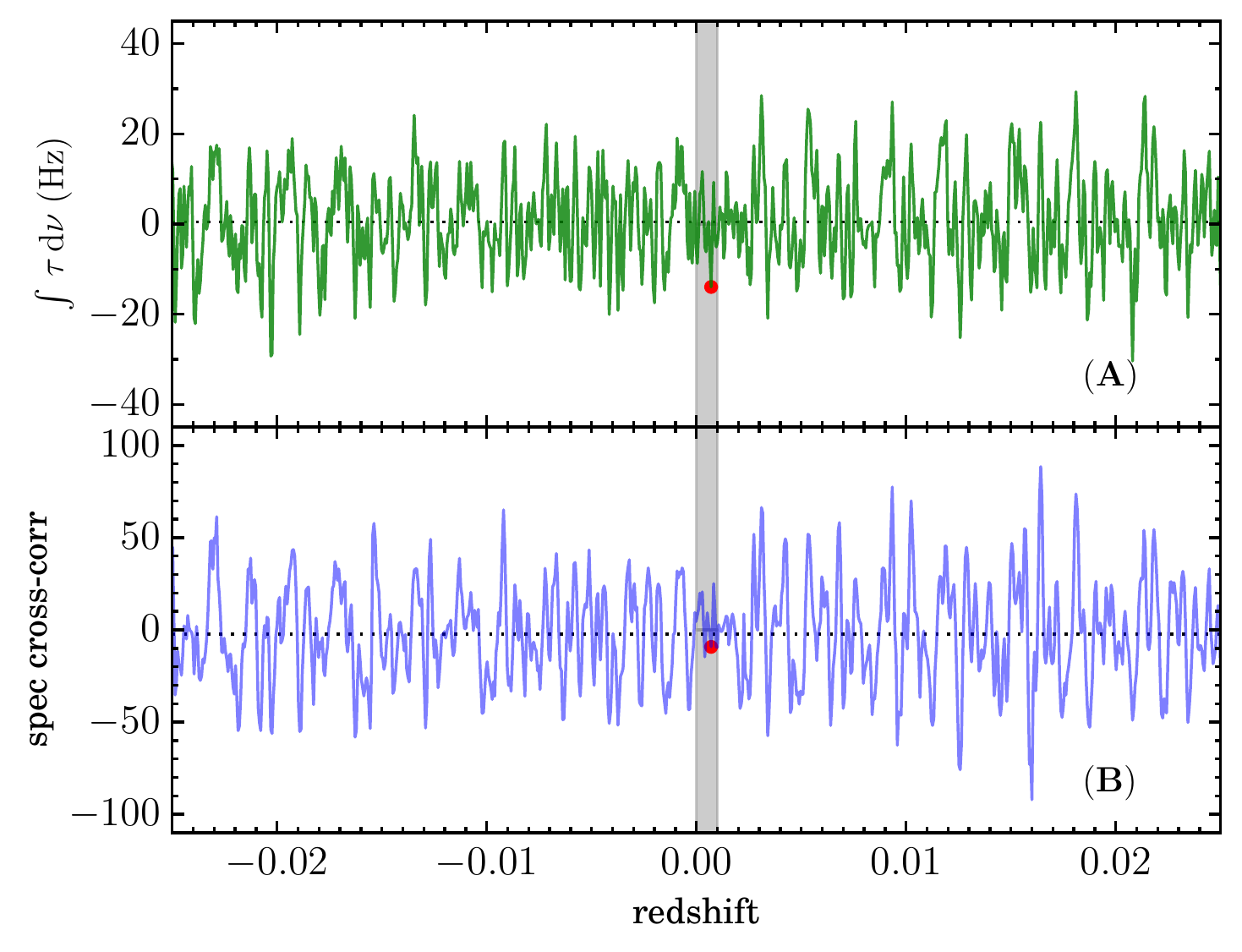}
    \includegraphics[width=0.48\textwidth]{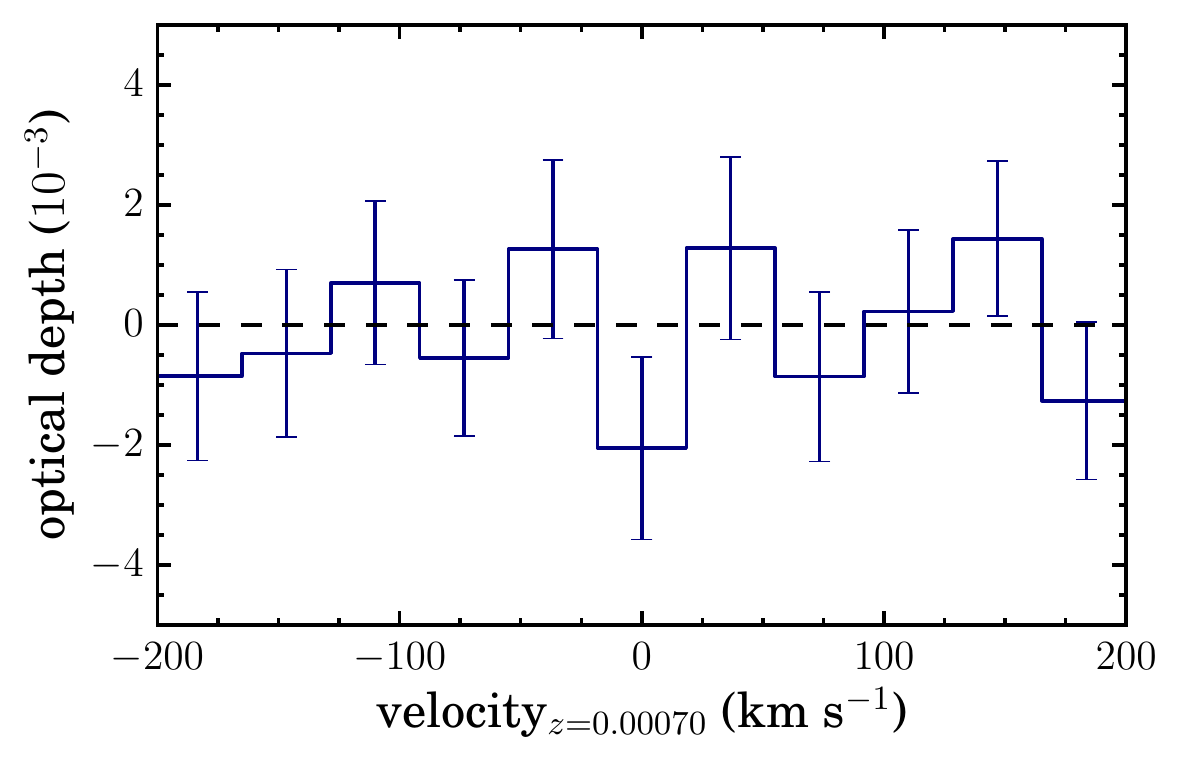}
    \caption{Our results for M 82. {\it Left:} Same as in Figure \ref{fig:casa_epsilon}, except applied to the C$\alpha$-transitions of M 82. We did not proceed with the stack cross-correlation because no significant outlier was identified in (B). The shaded region in gray shows the redshift range probed by \citetalias{Morabito2014}. The red data point shows marks $z_{\mathrm{M}82} = 0.00070$, the redshift corresponding to the peak of RRL emission, as reported by \citetalias{Morabito2014}. {\it Right:} The stacked spectrum of M 82, in velocity units, with respect to $z_{\mathrm{M}82} = 0.00070$. The error bars reflect the standard deviation of a weighted mean (see Sec. \ref{sec:stackproc})}
    \label{fig:m82_rp}
\end{figure*}

M 82 is a prototypical starburst galaxy located $\sim$3.5 Mpc away \citep{Jacobs2009}, with a systemic velocity of $v_{\rm sys} = 209 \pm 4$ \kms\ \citep{Kerr1986}. It was observed with the LOFAR LBA and reported to have CRRLs in absorption \citepalias{Morabito2014} centered at $v_{\rm LSR} = 211.3^{+0.7}_{-0.5} $ \kms\ or $z_{\rm cen} = 0.00070$. The feature has a peak optical depth of $2.8^{+0.12}_{-0.10} \times 10^{-3}$ and  a FWHM  of $30.6^{+2.3}_{-1.0}$ \kms providing an integrated optical depth of $\int \tau {\rm d \nu} = 21.3$ Hz; errors were derived from the Gaussian fit. We were motivated to test our methods on the M 82 data as it makes for the most direct comparison to the 3C 190 spectra, where the spectral feature is narrow, has low signal-to-noise and the fraction of line-blanked channels to line-free channels ($\gtrsim10\%$) starts to induce non-negligible effects.

We used the spectra extracted from the 50 MHz -- 64 MHz observations of an unresolved M 82 (Morabito, priv. comm., \citetalias{Morabito2014}). The data have a channel width of 6.10 kHz, which corresponds to 28.6 \kms\ -- 36.7 \kms\ over this frequency range. We Doppler corrected the spectra to the Barycentric frame. During stacking, we interpolated each spectrum to a fixed velocity grid with a channel resolution of 36.7 \kms. Before implementing the flagging and stacking routine described in Section \ref{sec:stackproc}, we flagged two channels at the starting edge (lower in frequency) of the subband and one channel at the ending edge (higher in frequency). We cover the redshift ranges $-0.025 \leq z \leq 0.025$ in stacking, sampled at redshift intervals of $\delta z_{\rm sample} = 6 \times 10^{-5}$. We considered line-blanking regions of $\pm25$ \kms, $\pm50$ \kms, and $\pm100$ \kms. 

We did not find a significant outlier in the integrated optical depth across redshift or in the spectral cross-correlation (Sec. \ref{sec:spec_cc}) using any of the line-blanking widths.  An example, with a line-blanking region of $\pm 50$ \kms, is shown in Figure \ref{fig:m82_rp}. The redshift at which RRLs were reported in \citetalias{Morabito2014} is shown in the plots with a red, circular data point. Since we could not identify an outlier in the spectral cross-correlation, we did not proceed with the stack cross-correlation (Sec. \ref{sec:stack_cc}). 

In Figure \ref{fig:m82_rp} we also show the CRRL stacked spectrum centered at $z=0.00070$. The optical depth in the central channel is $(2.1 \pm 1.5) \times 10^{-3}$. The standard deviation in the optical depth of channels within $\pm 200$ \kms\ when excluding the central channel is $\sigma = 9.7 \times 10^{-4}$. This noise (which is an rms per channel) is about 1.5 times lower than the error attributed to the central channel. This mismatch arises from (1) interpolation and (2) non-uniform coverage across the channels. We compared the noise in the subband spectra before and after interpolation, and found that interpolation reduced the noise by 20 -- 30 percent; an overall 6 percent was due to an effective averaging of higher resolution channels to the coarse grid. This affects all channels of the spectrum. Furthermore, the number of data points averaged in channels which are line free is 20 -- 30 percent higher than in the line channels; this results in a lower rms in the line-free channels only. Accounting for these two effects results in an rms per channel of $\sigma = 1.4 \times 10^{-3}$, a peak optical depth of $(2.5 \pm 1.4) \times 10^{-3}$, and for an effective frequency of 55.5 MHz, an integrated optical depth of $\int \tau {\rm d \nu} = 18 \pm 10$ Hz.

The properties of the spectral feature that we find centered at $z=0.00070$ are consistent with the values reported by \citetalias{Morabito2014}. However, the significance relative to the noise is 1.8$\sigma$. When we compare the value of the integrated optical depth at $z=0.00070$ with values obtained for all of the other redshifts tested (see Figure \ref{fig:m82_rp}), its value is 1.5 times the standard deviation. 

Moreover, the value of the spectral cross-correlation at $z=0.00070$ does not appear significant in comparison to the values across the full range of redshifts (0.35 times the standard deviation of the cross-correlation values), nor in the redshift range probed by \citetalias{Morabito2014} (z = 0.0 -- 0.001; see gray, shaded region in Figure \ref{fig:m82_rp}).  The latter may be unexpected, given that \citetalias{Morabito2014} found a peak in the cross-correlation when the stack was centered on a redshift of $z=0.00073$. We note that the spectral cross-correlation we have implemented differs from the \citetalias{Morabito2014} approach in three ways, which we explain in the following paragraphs.

Firstly, a separate fit and subtraction of the continuum is done at each redshift. It is  essential to include this step because, by definition, residuals of the fit have been minimized in the line-free channels; thus any spectral search in the ``off'' redshifts, the line-free channels, will be consistent with noise. Furthermore, the error within the line-blanked channels will be amplified since the continuum has not been directly estimated from those channels. This is especially true when the number of channels available to estimate the continuum (about 25 channels) is small \citep{Sault1994}. 

Secondly, we generate a new template spectrum for each redshift. The number of spectral lines available for stacking varies across redshift. For example, there are 8 fewer spectral lines available to stack at $z= 0$ compared with $z=0.00070$. The cross-correlation would naively appear lower at $z=0$ only because fewer lines fall within a searchable region. 

This leads to the third difference, which is that we have normalized the cross-correlation value by the number of spectral lines stacked (or equivalently, by the area of the cross-correlation function). Without a normalization, the cross-correlation values are not directly comparable for functions of different areas. 

We went about reproducing the \citetalias{Morabito2014} results following their methods, in order to understand the limitations of our method and verify the results. The major change in the procedures implemented to obtain the line profile involves smoothing the combined subband spectra with a filter rather than interpolating the subband spectra to a fixed velocity grid and averaging. The minor changes involved in flagging were also included. We reproduced the spectrum of \citetalias{Morabito2014} and investigated the noise properties, the line properties, and the distribution in those values when stacked across redshift. We describe the procedure and discuss the results in the Appendix \ref{append:m82}. The spectrum we obtained has line properties consistent with those of \citetalias{Morabito2014}. However, with an updated noise estimate (Figure \ref{fig:m82_rl_spec}), the feature has a 2.2$\sigma$ strength.

We conclude that deeper LOFAR observations will be needed to investigate CRRLs in the spectrum of M 82. Since the line-width of the reported feature is less than a channel width, the peak strength would be underestimated and width overestimated; processing the data at higher spectral resolution will also be necessary. A 3$\sigma$ upper limit on the integrated optical depth is 26.5 Hz for a line-width of 36 \kms.

\section{3C 190}
\label{sec:3c190}

\begin{figure*}
    \centering
    \includegraphics[width=0.45\textwidth]{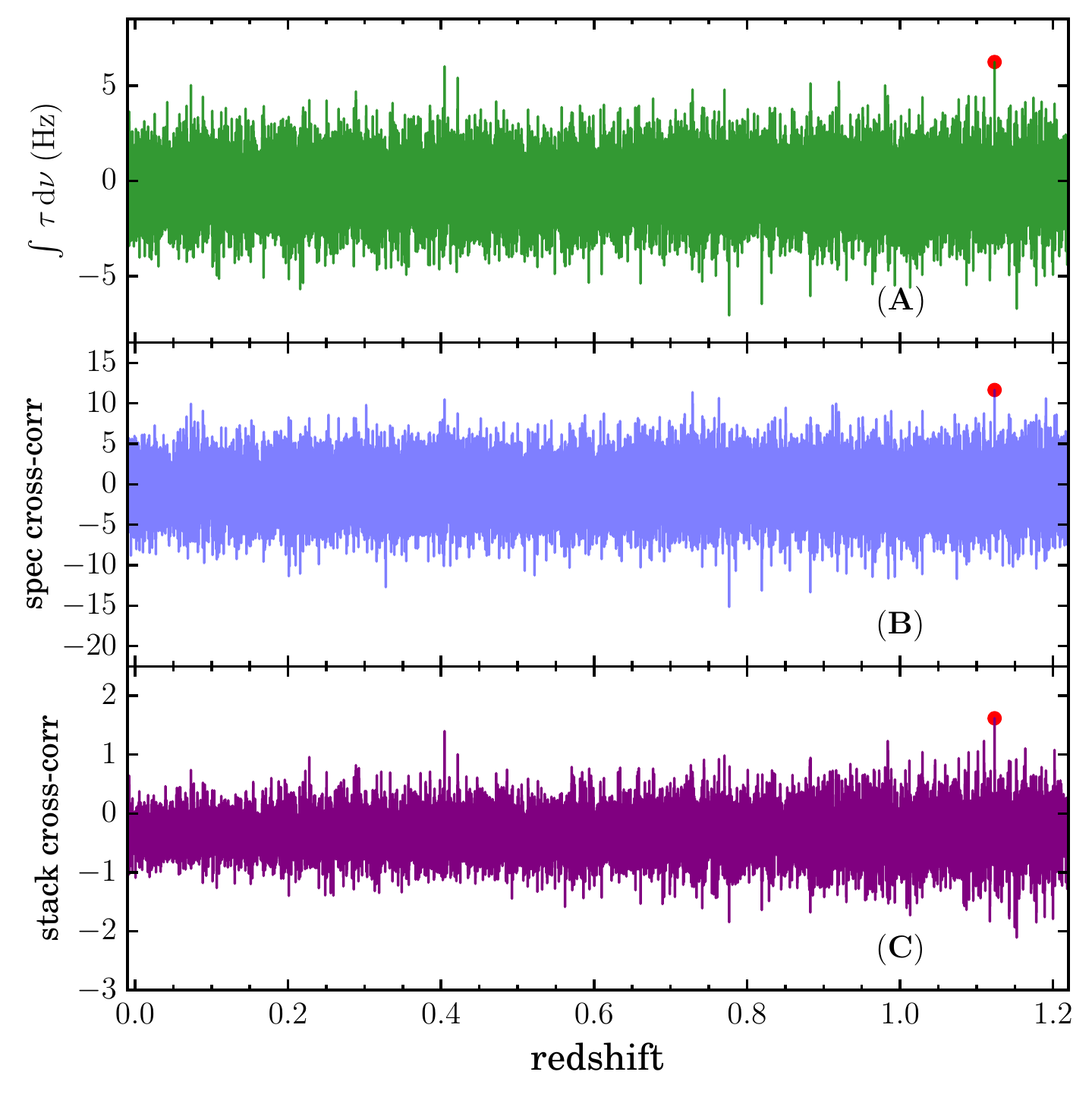}
    \includegraphics[width=0.5\textwidth]{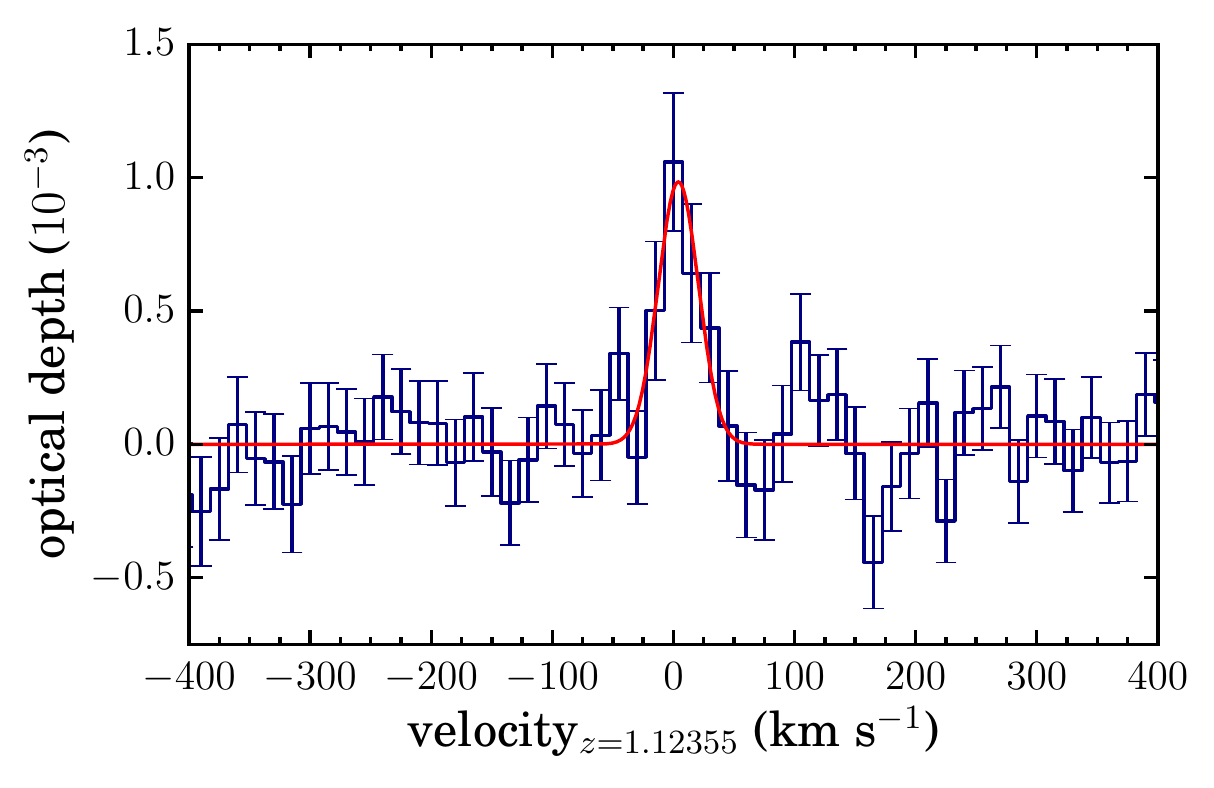}
    \caption{Our results for 3C 190. {\it Left:} Same as in Figure \ref{fig:casa_epsilon} except applied to the C$\alpha$-transitions of 3C 190. The red data point marks the significantly identified redshift of $z=1.12355$. {\it Right:} The spectrum resulting from stacking C$\alpha$-transitions at $z=1.12355$ in 3C 190, using a line-blanking region of $\pm 25$ \kms. }
    \label{fig:3c190_cc}
\end{figure*}

We identified 3C 190 as a candidate to search for RRLs at high redshift as it is a bright radio source at LOFAR frequencies, compact (4 arcsec in size), and has been detected via HI absorption \citep{Ishwara-Chandra2003} and Mg II absorption \citep{Stockton2001}, which are both indicative of cold gas. The data were Doppler corrected to the Barycentric rest frame, and the stacked spectra were interpolated to a velocity resolution of 15 \kms . We searched the redshift ranges $-0.01 < z < 1.22$ and considering line-blanking regions of $\pm7.5$ \kms, $\pm30$ \kms, $\pm45$ \kms, and $\pm60$ \kms.

We find one significant outlier with the line-blanking region of $\pm7.5$ \kms . However, no outliers were found with other line-blanking regions. The results of the method when line-blanking $\pm7.5$ \kms\ are shown in Figure \ref{fig:3c190_cc}. We ultimately find $\alpha$-transition RRLs at $z=1.12355$ assuming carbon, as reported in \cite{Emig2019a}. In  Figure \ref{fig:3c190_cc}, we also show the spectrum when extending the line-blanking region to $\pm25$ \kms as in \cite{Emig2019a}. We do not find a significant signal at the systemic redshift of 3C 190 or at the redshifts of the reported absorption features and place a $3\sigma$ upper limit of 4.6 Hz for a line width of $\pm7.5$ \kms. If the noise scaled in proportion to the line-blanking regions, we would expect an upper limit of 18.6 Hz for a line width of $\pm45$ \kms, however given the distribution of integrated optical depth, we find a $3\sigma$ upper limit of 22.7 Hz is more representative. This indicates that systematics such as the poly-phase filter residuals are present in the data, and noise is being amplified within the line-blanked region due to improper continuum subtraction. We find the noise to be amplified by a factor of 1.22 when roughly 4 of 26 channels are blanked.

\section{Discussion of Methods} 
\label{sec:discuss}

The data reduction strategy we have laid out covers basic calibration with an improvement on the bandpass calibration, which is of particular importance for RRL observations. Future strategies that include corrections for ionospheric effects, a phase offset present in some stations, as well as including stations with longer baselines to increase spatial resolution could be beneficial to the strategy here \citep{VanWeeren2016, Williams2016}. Although with the set up and configuration processed here, these are not dominant effects. Prefactor\footnote{https://github.com/lofar-astron/prefactor.git} 3.0 would enable this and has been designed with spectroscopic studies in mind \citep{DeGasperin2019}.

We have presented a three step method to search for radio recombination lines most pertinently in a low signal-to-noise regime. This method can be applied to observations of any telescope with sufficiently large fractional bandwidth that allow for spectral stacking. This includes recombination lines at higher frequencies --- for example, in the L band, where 17 $\alpha$-transitions lie between rest frequencies of 1--2 GHz. We have demonstrated our method in a variety of regimes. 

As a proof of concept, we showed the behavior of stacked recombination lines with high signal-to-noise (see Figures \ref{fig:alpha_itau} -- \ref{fig:alpha_ccstack}). When comparing the optical depth integrated within a fixed region for a range of input redshifts/velocities, we see the actual redshift is clearly the most prominent in comparison. However, since the lines are also detected with significance individually, redshifts corresponding to mirrors of the signal are of course, detected with significance as well. In this regime the spectral cross-correlation (Section \ref{sec:spec_cc}) identifies the actual redshift most uniquely, whereas the stack cross-correlation (Section \ref{sec:stack_cc}) is more sensitive to the stack mirrors and therefore the relative significance of the mirrors is enhanced.

We also show and examine low signal-to-noise features, and we do so in two regimes: broad features and narrow features. Broad features can be defined such that their line width is greater than the difference in frequency spacing between adjacent lines of principal quantum number $\sf{n}$, $\Delta\nu_{\mathrm{FWHM}} \gtrsim (\Delta\nu_{\mathsf{n-1, n}} - \Delta\nu_{\mathsf{n, n+1}})$. In this case, at redshifts corresponding to the mirrors of the stack, the slight misalignment in frequency/velocity space fall within the FWHM of the line, causing the ratio between the integrated signal and the mirrors to be closer to unity. An example of this regime is the $\epsilon$-transition search, were the line width is $\Delta\nu_{\rm FWHM}=6.4$ kHz and $(\Delta\nu_{\mathsf{n-1, n}} - \Delta\nu_{\mathsf{n, n+1}}) = 1$ kHz for $\mathsf{n} = 801$. 
Conversely, narrow features we define as $\Delta\nu_{\mathrm{FWHM}} \lesssim (\Delta\nu_{\mathsf{n-1, n}} - \Delta\nu_{\mathsf{n, n+1}})$, where the narrowness of the line causes minimal resonance in the mirrors, and the ratio of the integrated optical depth between $z_{cen}$ and $z_{mirror}$ approaches $1/N$ with $N$ being the number of lines stacked. An example of this is H$\alpha$, where $\Delta \nu_{\mathrm{FWHM}} = 0.7$ kHz and $(\Delta\nu_{\mathsf{n-1, n}} - \Delta\nu_{\mathsf{n, n+1}}) = 4.0$ kHz for $\mathsf{n} = 468$.

When stacking broad spectral lines, the integrated optical depth at the redshift of the stack mirrors is maximally enhanced. Therefore, the stack mirrors may be clearly distinguished (see Figure \ref{fig:casa_epsilon}) and the ratio between the integrated signal at the actual redshift and at the mirror redshifts is less extreme. In the presence of well behaved bandpasses and high $N$ statistics, the integrated optical depth and spectral cross-correlation may be sufficient to identify the feature. However, the stack cross-correlation results in a distribution of redshifts, with uncertainties of $\delta z = \Delta \nu_{\mathsf{n, n+1}} / \nu_{\mathsf{n}} $. When, for example, bandpass estimation is distorted at ``off'' redshifts (by the very presence of the spectral lines) or in the case of low $N$ statistics, additional scatter in the integrated optical depth could result and the use of the stack cross-correlation becomes essential.

On the other hand, narrow features produce only a low level mirror of the stack at flanking redshifts, and in all three steps of the method, only the actual redshift of the source (no mirror redshifts) appear as outliers (see Figure~\ref{fig:casa_h_c0_alpha}). In this regime, the stack cross-correlation becomes the most sensitive probe. 

Useful tips that we have learned using our method include (1) the significance of the feature is maximized in each step of the procedure when the optical depth is integrated within a region roughly half the size of the line-blanking region; (2) it is essential to minimize residuals of high signal-to-noise features in order to reliably probe additional low signal-to-noise features; and, (3) the stack cross-correlation is most effective when only one mirror is included in the cross-correlation function. 

Focusing briefly on the implications of integrated optical depth distributions, we find that the apparent amplification of noise shown in the distribution of integrated optical depth in M 82 and 3C 190 indicates that the noise in the stacks is not purely Gaussian; several effects may be at play. Systematics may be present in the spectra. The continuum removal may amplify noise within the line-blanked region, with too few channels to estimate the continuum. Lastly, fluctuations due to low $N$ statistics and non-uniform coverage across channels may be contributing to additional scatter. For example, in Cas A the number of line-free channels is large ($\sim200$) and the fraction of line to continuum channels is typically $\leq 0.01$, resulting in a negligible error of estimating the continuum within the line-region. The fraction of velocity channels in which the coverage is uniform is also larger. Indeed, across redshift, we find that the scatter in the peak optical depth of the stack matches the median channel noise. 

Nonetheless, the detection of an RRL in the first AGN we searched with LOFAR holds promise for future exploration. For typical low-frequency RRL optical depths of $10^{-3} - 10^{-4}$, it is reasonable and feasible to search in the 3CR sources \citep{Edge1959, Bennett1962} of the northern hemisphere, which have flux densities  $>9$ Jy at 178 MHz. With 328 sources in the northern hemisphere and an average source density of $1.6 \times 10^{-2}$ deg$^{-2}$, about 0.8 3CR sources will fall in a $\sim$50 deg$^{2}$ pointing of the LOFAR Two Meter Sky Survey \citep{Shimwell2019}. In light of our results above, it will be necessary to perform a continuum estimation and subtraction across multiple subbands, for the survey frequency resolution of 16 channels per subband, to reach adequate sensitivities.

\section{Conclusion} 
\label{sec:conclude}

RRLs are largely unexplored at low frequencies although they uniquely can provide long sought after physical conditions of the cold-neutral (and warm-ionized) phase(s) of the ISM. We have described methods to calibrate and extract RRLs in low-frequency ($<$ 170 MHz) spectra. Starting with LOFAR observations that are optimized for extragalactic sources (where line widths may be 10 -- 100 \kms\ and can plausibly be probed to $z \sim 4$), we discussed spectroscopic data reduction. 
We then showed a procedure in which spectra are stacked and cross-correlated to identify low signal-to-noise features. One cross-correlation is taken between a template spectrum and the observed spectrum, both in optical depth units, where the location of each line is utilized. A second cross-correlation incorporates the average spacing between lines (and their line width);  the integrated optical depth over a range of redshifts is cross-correlated with the distribution of the template spectrum over a range of redshifts, corroborating what we refer to as ``mirrors'' of the stack at flanking redshifts. 

Our method was developed to search blindly in redshift for RRLs in the LOFAR HBA spectrum of 3C 190, in which we have identified an RRL in emission at $z=1.12355 \pm 0.00005$ (assuming a carbon origin). This was the first detection of RRLs outside of the local universe \citep{Emig2019a}. To demonstrate and test the limitations of the method, we also apply it to existing LOFAR observations of the sources Cas A and M 82. 

We have re-analyzed the 55 -- 78 MHz LOFAR spectra of Cas A. Using our methods, we discover three new detections in the data, plus the original detections of \cite{Oonk2017}. We significantly detect C$\alpha$($\mathsf{n}$=467), C$\beta$(590), C$\gamma$(676), C$\delta$(743), and C$\epsilon$(801) transitions associated with the line-of-sight -47 \kms\ and/or -38 \kms\ components. 
This is the first detection of an $\epsilon$-transition ($\Delta {\sf n} = 5$) at low radio frequencies. We also find H$\alpha$(468) in emission at 64.08 MHz with $\int \tau {\rm d}\nu = (-0.21 \pm 0.04)$ Hz and a FWHM of 3.1 \kms\  resulting in one of the lowest frequency and most narrow detections of hydrogen. The line-width directly associates this hydrogen with the cold, ionized component of the ISM; this is further supported by our updated modeling of the gas physical properties with best fit conditions of $T_e = 40$ K, $n_e = 0.06$ cm$^{-3}$, and $EM_{H+} = 0.0012$ pc cm$^{-6}$.  Additionally, we detect C$\alpha$ associated with the Orion Arm at 0 \kms at these frequencies for the first time. 

For the 55 -- 64 MHz spectra of the nearby starburst galaxy M 82, we recover the line properties reported by \cite{Morabito2014} and find the integrated optical depth to be  $\sim$2$\sigma$ relative to the noise. A $3\sigma$ upper limit on the integrated optical depth is 26.5 Hz. Follow-up LOFAR observations reaching deeper sensitivities and higher spectral resolution will be worthwhile.

We find that LOFAR observations using 32 channels per subband is not optimal for RRL studies. Because the number of channels available to estimate the continuum is low, the noise in the line blanking region is amplified. Currently continuum subtraction can only be estimated within a single subband. As the non over-lapping nature of the narrow (195.3 kHz) subbands makes a smooth bandpass calibration difficult, future observing bands with larger contiguous frequency coverage would enable deeper searches of RRLs in extragalactic sources. 

\begin{acknowledgements}

The authors would to thank Leah Morabito for providing the LOFAR LBA spectra of M 82 and for the discussions, and Reinout van Weeren for guidance and careful review of the manuscript.

KLE, PS, JBRO, HJAR and AGGMT acknowledge financial support from the Dutch Science Organization (NWO) through TOP grant 614.001.351. AGGMT acknowledges support through the Spinoza premier of the NWO. MCT acknowledges financial support from the NWO through funding of Allegro. FdG is supported by the VENI research programme with project number 639.041.542, which is financed by the NWO. Part of this work was carried out on the Dutch national e-infrastructure with the support of the SURF Cooperative through grant e-infra 160022 \& 160152. 

This paper is based (in part) on results obtained with International LOFAR Telescope (ILT) equipment under project codes \texttt{LC7\_027, DDT002}. LOFAR \citep{vanHaarlem2013} is the Low Frequency Array designed and constructed by ASTRON. It has observing, data processing, and data storage facilities in several countries, that are owned by various parties (each with their own funding sources), and that are collectively operated by the ILT foundation under a joint scientific policy. The ILT resources have benefited from the following recent major funding sources: CNRS-INSU, Observatoire de Paris and Universite d'Orleans, France; BMBF, MIWF-NRW, MPG, Germany; Science Foundation Ireland (SFI), Department of Business, Enterprise and Innovation (DBEI), Ireland; NWO, The Netherlands; The Science and Technology Facilities Council, UK; Ministry of Science and Higher Education, Poland.

\end{acknowledgements}

\bibliographystyle{aa}
\bibliography{papers-thesis2}

\begin{appendix}

\section{Subband spectra of Cas A}
\label{append:casa_spec}

In Figure \ref{fig:casa_spec}, we show Cas A spectra for which the continuum has been removed, in units of optical depth as a function of frequency. These spectra are shown prior to stacking in order to illustrate typical properties of the observations. For example, each subplot shows the spectrum of a single subband and the channels remaining unflagged out of the original 512 channels per subband. How often spectral lines of the various transitions -- $\alpha$, $\beta$, $\gamma$, $\delta$, $\epsilon$ -- fall and how often they overlap can be grasped. In addition, the channels used to estimate the continuum are those outside of the shaded regions. 

\begin{figure*}
\centering
\includegraphics[width=0.92\textwidth]{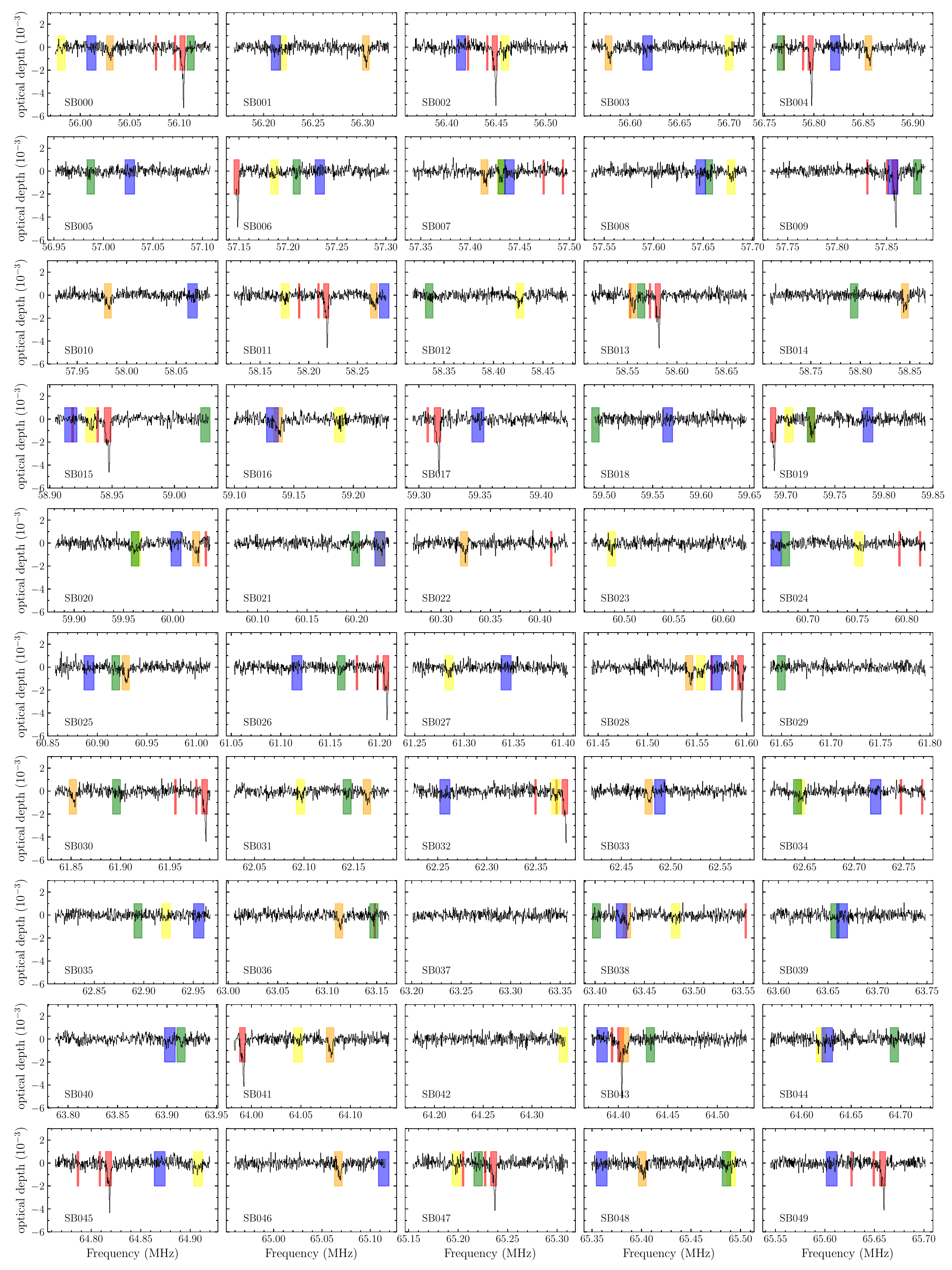}
\caption{ {\it (cont.)}}
\label{fig:casa_spec}
\end{figure*}

\begin{figure*}
\ContinuedFloat
\centering
\includegraphics[width=0.92\textwidth]{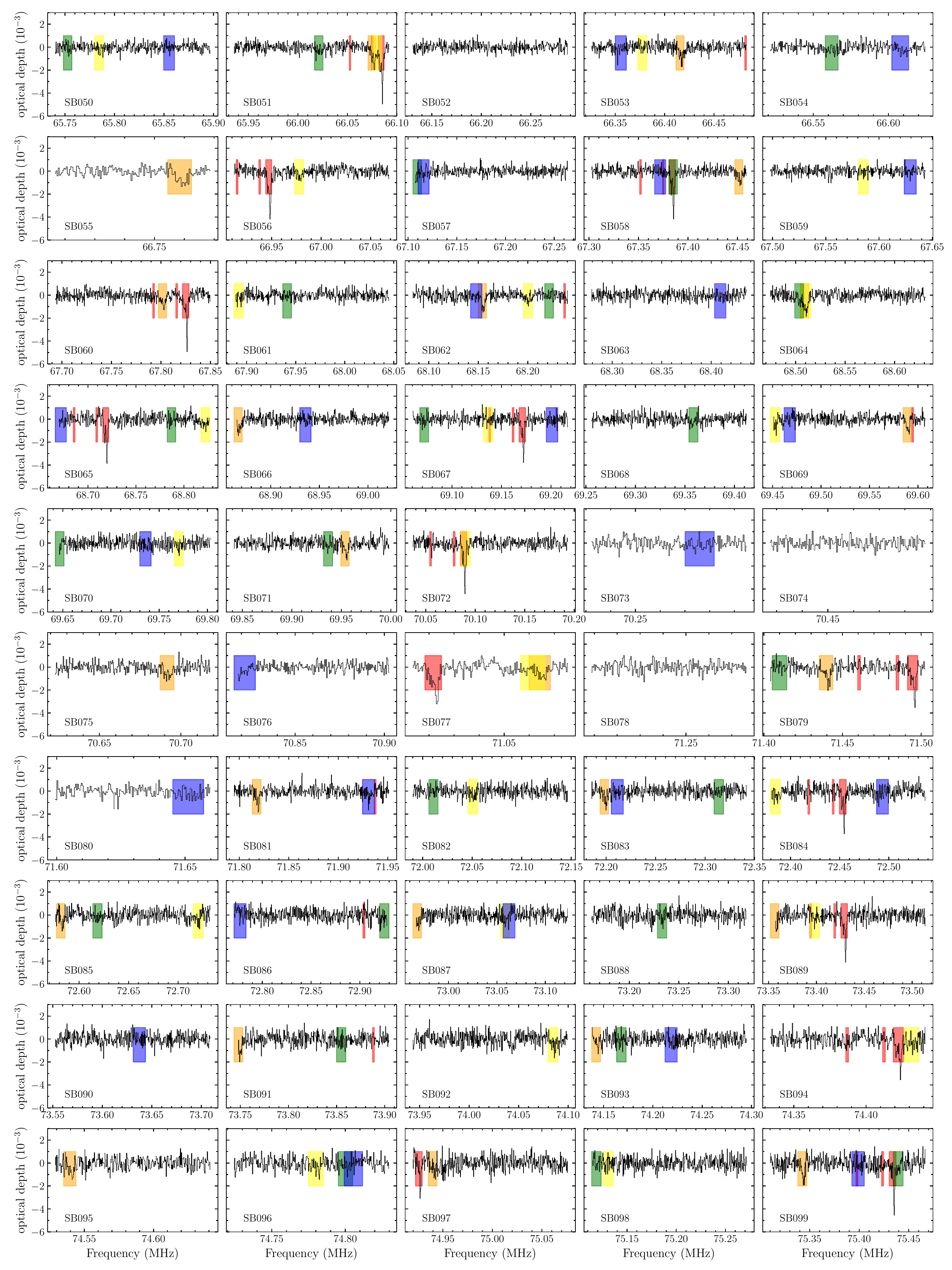}
\caption{ {\it (cont.)} }
\end{figure*}

\begin{figure*}
\ContinuedFloat
\centering
\includegraphics[width=0.92\textwidth]{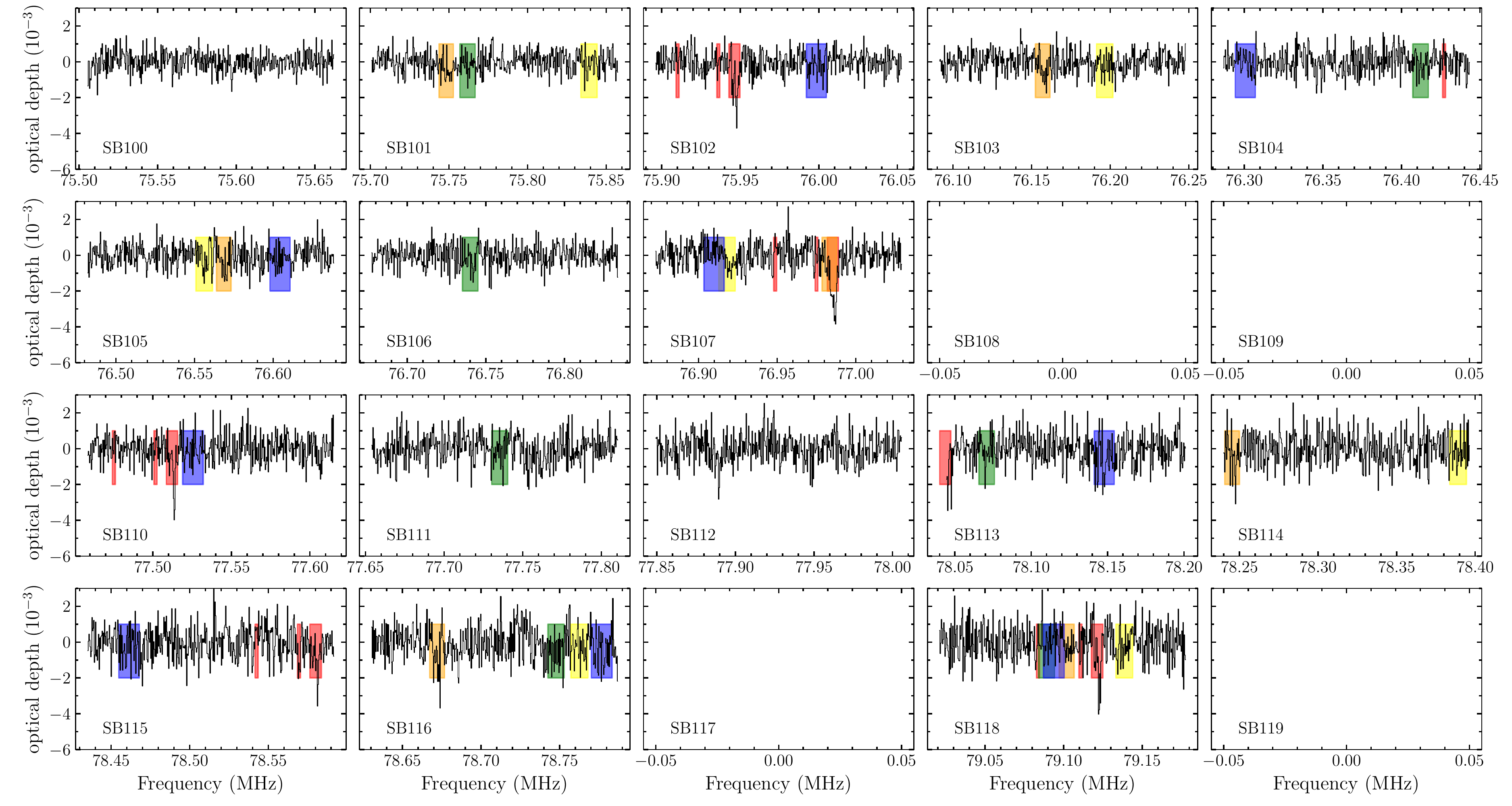}
\caption{ The subband spectra of Cas A, shown after continuum removal, as the optical depth as a function of frequency. The red shaded regions cover the location of $\alpha$-transitions, orange regions cover $\beta$-transitions, yellow regions cover $\gamma$-transitions, green regions cover $\delta$-transitions, and blue regions cover $\epsilon$-transitions. We note that the widest shaded regions in red cover the spectral lines of the -47 km/s and -38 km/s C$\alpha$ components; two recurring thin shaded regions in red represent the weak, narrow spectral lines of the 0 \kms\ C$\alpha$ component and H$\alpha$ at -47 \kms.}
\end{figure*}

\section{Spectral properties of M 82 applying M14 criteria}
\label{append:m82}

In this section we first discuss the criteria implemented in \citetalias{Morabito2014} and compare it with the procedure generally used in this paper (Section~\ref{sec:m82}). We reproduce the spectrum of \citetalias{Morabito2014} and analyze the noise properties for the same assumed redshift. Finally, using the \citetalias{Morabito2014} flagging and processing, we implement our three step method to search for a significant outlier.

The stacking procedure used by \citetalias{Morabito2014} differs from our own. Instead of interpolating the spectra to a fixed velocity grid and averaging, they combine the unique velocity samples of the data and smooth the values with a Savitzky-Golav filter \citep{Savitzky1964}. This filtering has the superior advantages that when tuned properly, the peak and width line properties are well preserved \citep{Savitzky1964, Bromba1981, Press1990}. On the other hand, it requires that data be uniformly sampled, which is not the case for our stacked RRL spectra, and since the filter-width requires precise tuning \citep{Bromba1981}, it is more cumbersome for our aim of performing large-scale systematic searches. 

Next, we replicated the \citetalias{Morabito2014} results, with the procedures outline in that paper. We computed a Doppler correction to the barycentric rest-frame and subtracted $v_{\rm Doppler} = 10.24$ \kms . We flagged five subbands with large rms. A redshift of $z=0.00073$ was chosen by \citetalias{Morabito2014} because a peak in their spectral cross-correlation was found at this redshift and it maximized the signal-to-noise of the $z=0.00070$ feature. Thus, we assumed $z=0.00073$ and compared the expected frequency of the CRRL $\alpha$-transitions with the frequencies of the subband channels. We discard all subbands that do not contain a spectral line and additionally remove subbands where the spectral line falls closer than six channels from the subband edge. Then the first three channels and the last three channels were flagged. At this point, 23 subbands remained for processing.

We then converted the spectra into velocity units. Comparing the line-blanking region of $\pm50$ \kms\ with the channel velocities, we found that all 23 subbands had at least one unflagged channel on either side of the line-blanking region. \citetalias{Morabito2014} note that one subband did not have at least one channel remaining between the line-blanking region and the unflagged channels on the subband edge, and thus that subband was discarded. To identify the additional subband flagged, we identified three subbands which had a line blanking region separated from the edge by one channel: SB107, SB120, SB168. We created stacked spectra (see below for stacking procedure) in which just one of the subbands was flagged. We fit a Gaussian to the final spectrum and compared the noise in the line free region. When SB168 was excluded, the peak of the Gaussian, the center velocity of the fit, the FWHM, and the noise matched closest with the values reported in \citetalias{Morabito2014}. Thus we additionally flagged SB168. 

We blanked channels within $\pm50$ \kms\ of the line center and used the line-free channels to make a fit to the continuum. Each subband contained 31 channels to start with, after 6 total were flagged, and typically 3 were line-blanked, this resulted in typically 22 channels available to fit the continuum. The order of the polynomial fit to the continuum was not chosen by eye (among a first or second order) as done in \citetalias{Morabito2014}; instead, a first and second order polynomial was fit, and the polynomial order which resulted in a lower $\chi^2$ was taken. We divided all channels by the fit and subtracted one, obtaining spectral units in terms of optical depth. The optical depth values of a subband were given a weight $ w = \sigma^{-1}$, where $\sigma$ is the standard deviation of line-free channels.

\begin{figure}
\includegraphics[width=0.48\textwidth]{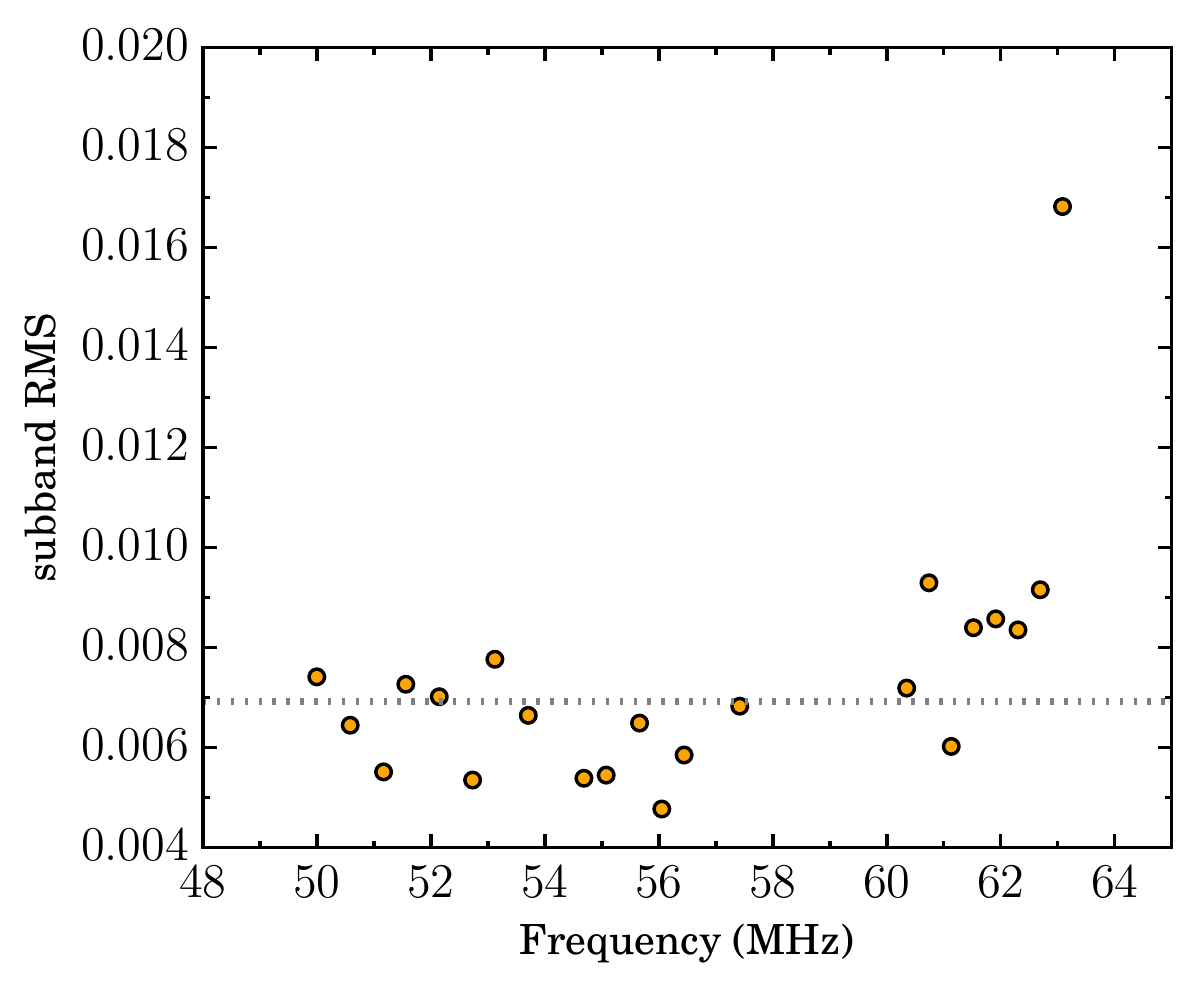}
\caption{The RMS of each M~82 subband plotted against the average frequency of the subband. The gray dotted lines represents the median value of $6.9 \times 10^{-3}$. These subbands were flagged and processed following the criteria of \citetalias{Morabito2014}. The RMS was determined from line-free channels only.}
\label{fig:noise_rl}
\end{figure}

\begin{figure}
\includegraphics[width=0.48\textwidth]{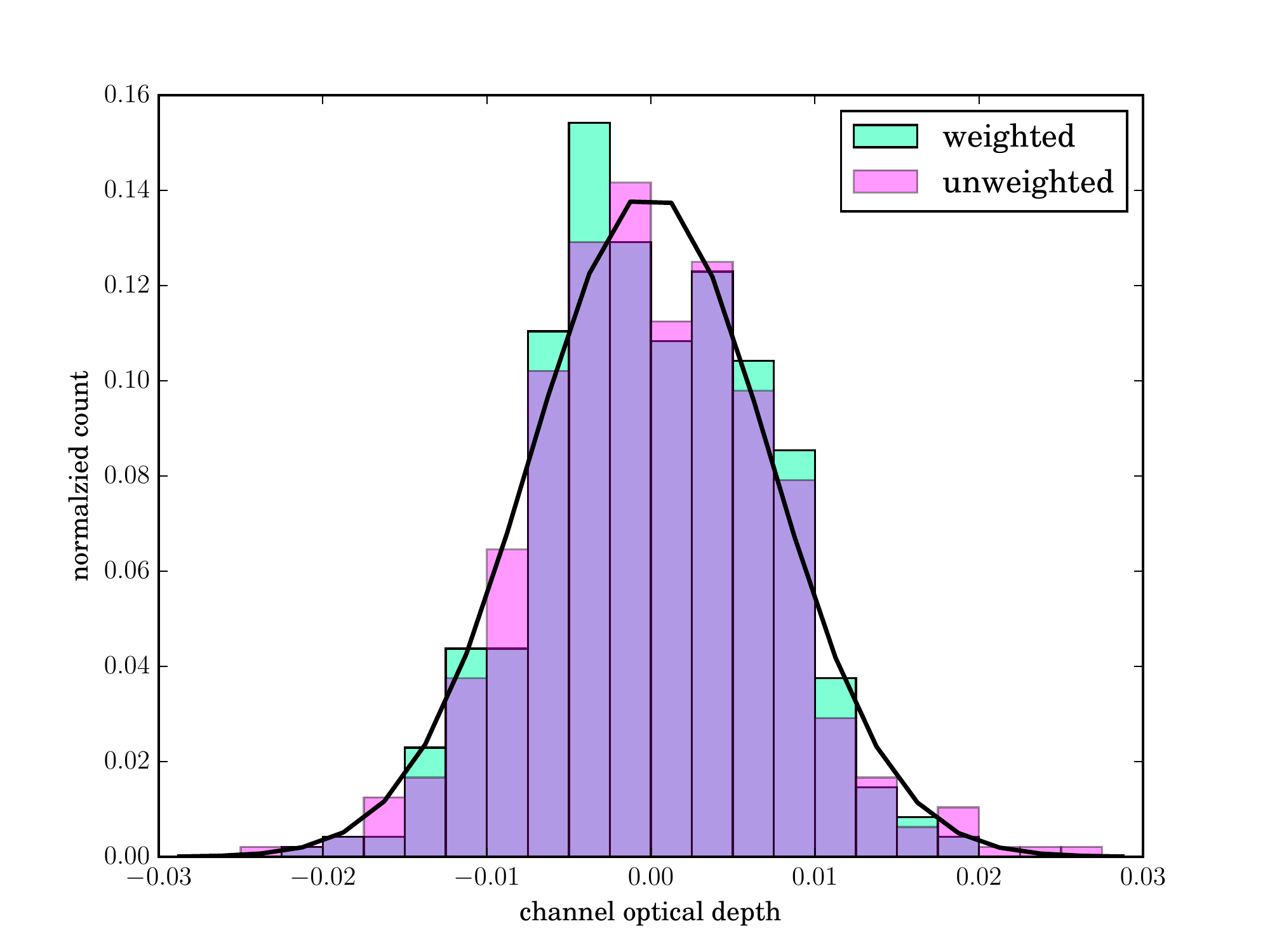}
\caption{Histogram of the optical depth of line-free channels, shown before (pink) and after (green) weighting with their overlap in purple. The black solid line shows the Gaussian fit to the weighted subbands. Unweighted, the fit to the distribution has a mean of $\mu_{uw} = -1.4 \times 10^{-3}$ and a standard deviation of $\sigma_{uw} = 7.2 \times 10^{-3}$. Weighted, the best fit Gaussian has properties of $\mu_{w} = -1.3 \times 10^{-3}$ and $\sigma_{w} = 7.3 \times 10^{-3}$. These subbands of M~82 were flagged and processed following the criteria of \citetalias{Morabito2014}. }
\label{fig:tauhist_rl}
\end{figure}

With these 22 subbands, we analyze the noise properties. We compute the noise of each subband by finding the standard deviation of the line-free channels. We plot this as a function of subband frequency in Figure \ref{fig:noise_rl}. The median value is $6.9 \times 10^{-3}$. We then collect the weighted optical depth values in all the line-free channels and plot them in a histogram (Figure \ref{fig:tauhist_rl}). The best fit Gaussian to the distribution has a mean value of $\mu = -1.4 \times 10^{-3}$ and a width of $\sigma = 7.3 \times 10^{-3}$. This shows that the uncertainty reflected in the data, before smoothing (or averaging) is  $7.3 \times 10^{-3}$. 

These noise properties contrast with \citetalias{Morabito2014}. In Figure 1 of \citetalias{Morabito2014}, the standard deviation of the optical depth is plotted as a function of subband frequency. The median value appears to be roughly $\sigma_{\rm median} \approx 1.5 \times 10^{-2}$.  However, the optical depth standard deviation of channels that have a velocity in the range of $ 50 {\rm km\, s}^{-1} < |v| \lesssim 150$ km s$^{-1}$ is quoted as $5 \times 10^{-3}$, prior to any smoothing of the data. In other words, \citetalias{Morabito2014} reported that the channels within the narrow velocity range they selected have a standard deviation that is a factor of three times lower than the standard deviation of the full sample of data. This indicates that the noise properties within that region do not reflect the actual uncertainty of the data.

\begin{figure}
\includegraphics[width=0.48\textwidth]{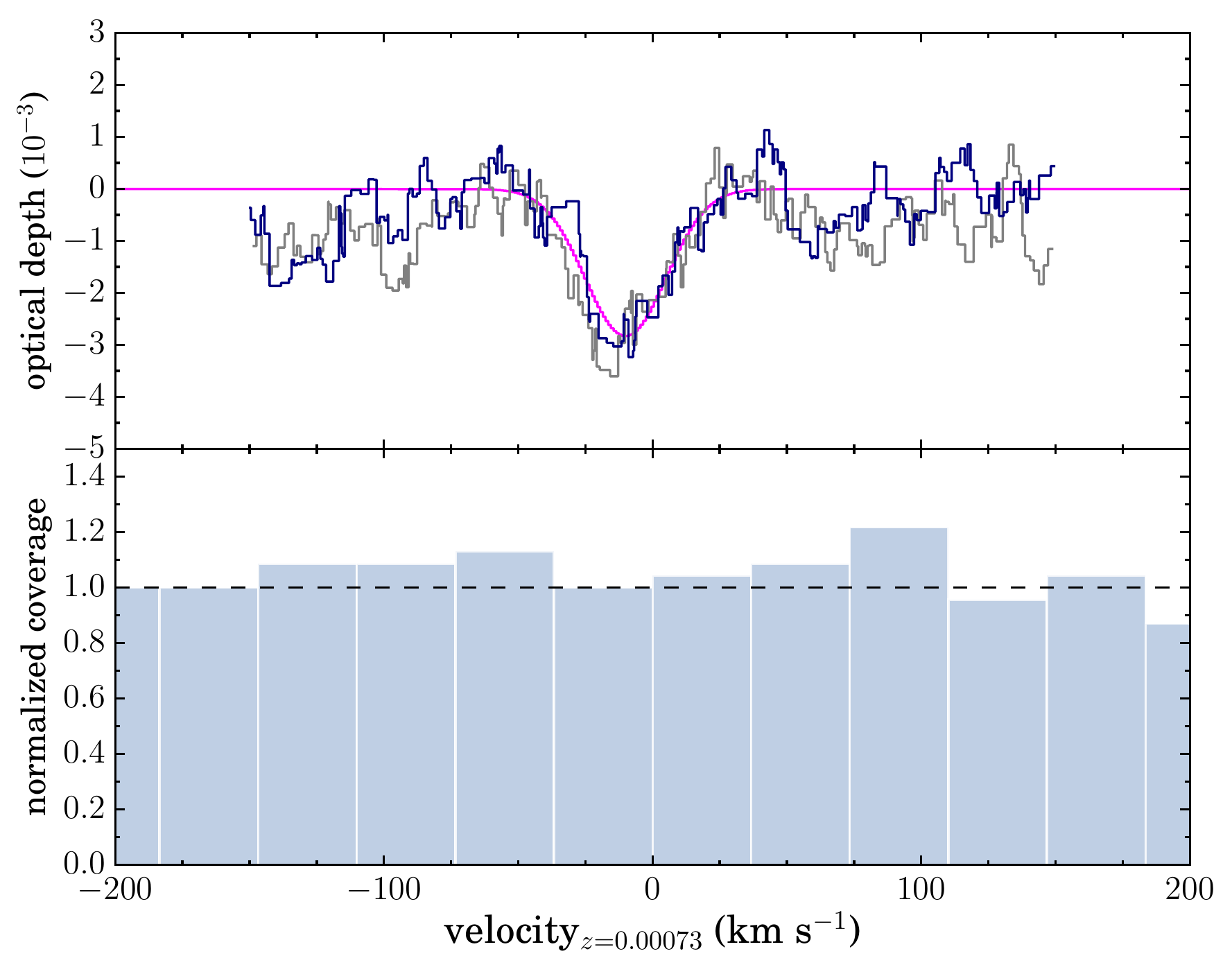}
\caption{The top plot shows the optical depth as a function of velocity for the RRL spectrum of M~82 stacked at $z_{\mathrm{M}82} =0.00073$ and smoothed with a Savitzky-Golay filter. In navy blue is the spectrum we reproduced following the criteria of \citetalias{Morabito2014}; the magenta line shows the best fit Gaussian. The standard deviation in the line-free channels is $6.5 \times 10^{-4}$. In gray is the final \citetalias{Morabito2014} spectrum (courtesy L. Morabito). The bottom plot shows the coverage, or number of subband data points, within velocity bins of 36.7 \kms. The coverage count has been normalized by 23, the number of data points in the bin where the line peak falls.} 
\label{fig:m82_rl_spec}
\end{figure}

We continued with the next step of processing in which the subband channels are aligned in velocity space and smoothed with the Savitzky-Golay filter. We show the results of Savitzky-Golay smoothing with a 31 data point filter width and a first order polynomial -- matching the properties which represent the main result of \citetalias{Morabito2014}. The spectrum we reproduce is shown in Figure \ref{fig:m82_rl_spec}, overplotted on the spectrum of \citetalias{Morabito2014} (courtesy L. Morabito). The properties of the spectral feature we reproduced are consistent within error. The best fit Gaussian (also shown in Figure~\ref{fig:m82_rl_spec}) to the feature that we reproduced has the following properties: a peak of $(-2.8 \pm 0.2) \times 10^{-3}$, a central velocity of $(-10.6 \pm 1.2)$ \kms, and a FWHM  of $(35.7 \pm 2.8)$ \kms. The optical depth standard deviation in the line-free channels of our spectrum is $\sigma = 6.5 \times 10^{-4}$.

A property of the Savitzky-Golay filter is that the noise is inversely proportional with $\sqrt{N}$ number of points of the filter width \citep{Savitzky1964}. Thus, we expect the noise ($7.3 \times 10^{-3}$) to be smoothed to roughly $1.3 \times 10^{-3}$ in our spectrum. However, this is a factor of about two greater than the standard deviation we measure ($6.5 \times 10^{-4}$). Incidentally, in the \citetalias{Morabito2014} analysis, spectra with a noise of $1.5 \times 10^{-2}$ are expected to have a noise of $2.7 \times 10^{-3}$ after smoothing. This is a factor of about 10 larger than the $3.3 \times 10^{-4}$ noise reported, and it is also consistent with the peak value of the RRL feature reported.

\begin{figure}
\includegraphics[width=0.48\textwidth]{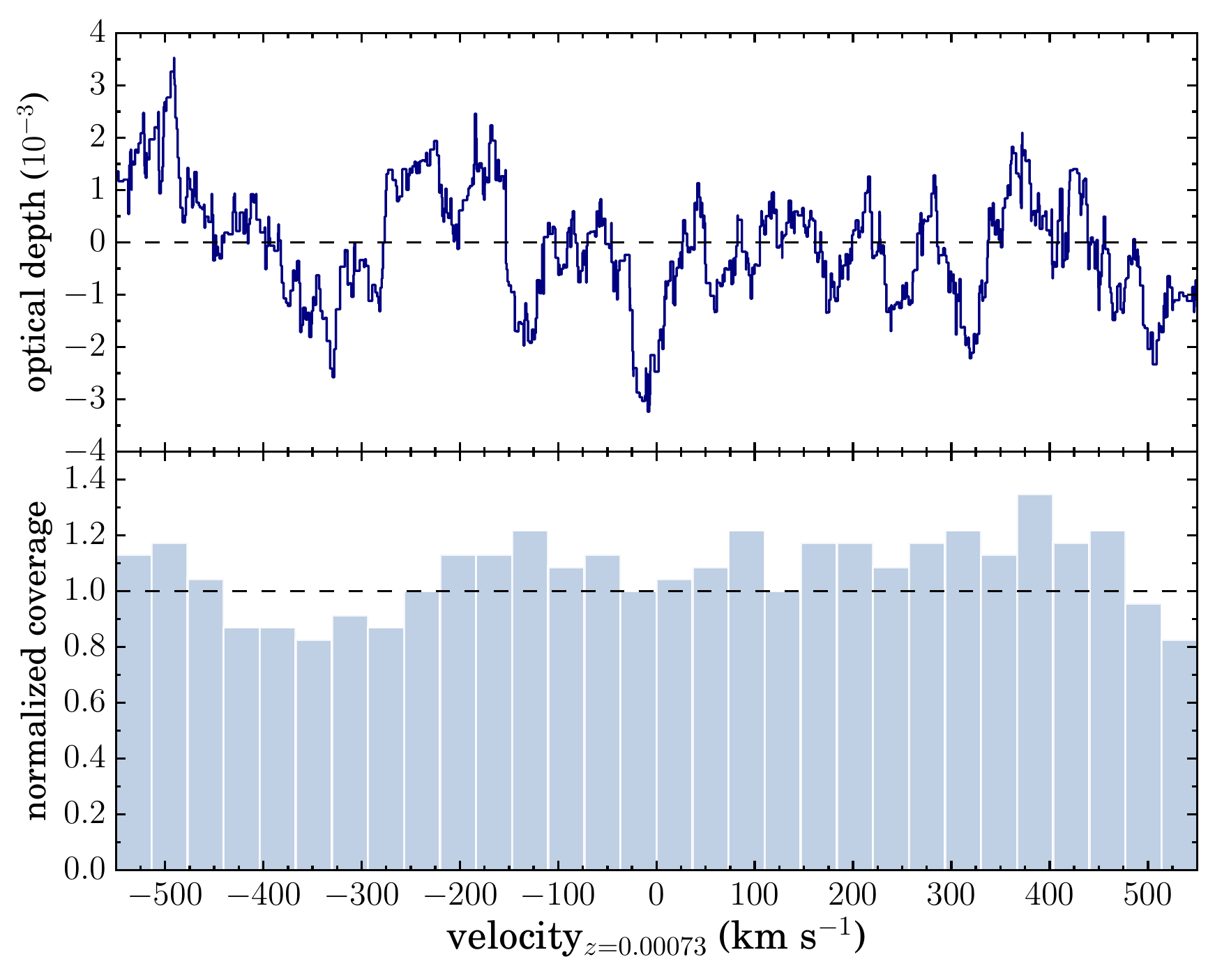}
\caption{Optical depth as a function of velocity, same as in Figure \ref{fig:m82_rl_spec}, except we have extended the coverage to larger velocities by including subbands of M~82 flanking those which contain a line. The standard deviation in the line-free channels after correcting for non-uniform coverage is $1.3 \times 10^{-3}$. }
\label{fig:m82_rl_covext}
\end{figure}

The narrow velocity range was selected by \citetalias{Morabito2014} because the coverage in velocity was approximately uniform (to within 20\%, e.g. see Figure \ref{fig:m82_rl_spec}). Here coverage refers to the number of subband data points that fall within a fixed velocity bin. However it is possible to extend the coverage to larger velocities by including flanking subbands which do not contain a spectral line. When including these subbands, we find the coverage continues to stay within 20\% (with a median value of an additional 13\%) out to $\pm550$ \kms.  We plot the spectrum which includes the additional subbands and its normalized coverage in Figure~\ref{fig:m82_rl_covext}. The noise we find in line-free channels out to $\pm 550$ \kms\ is $1.1 \times 10^{-3}$, which is now much closer to the expected noise of $1.3 \times 10^{-3}$ for completely uniform coverage. Quantifying the effects of applying the filter on non-uniformly sampled data is beyond the scope of this paper. Given this additional uncertainty and since the coverage is 13\% lower in the line region, we adopt the noise value of $1.3 \times 10^{-3}$. Thus, we find both the peak signal-to-noise and integrated optical depth equal to a 2.2$\sigma$ result. 

\begin{figure}
    \centering
    \includegraphics[width=0.48\textwidth]{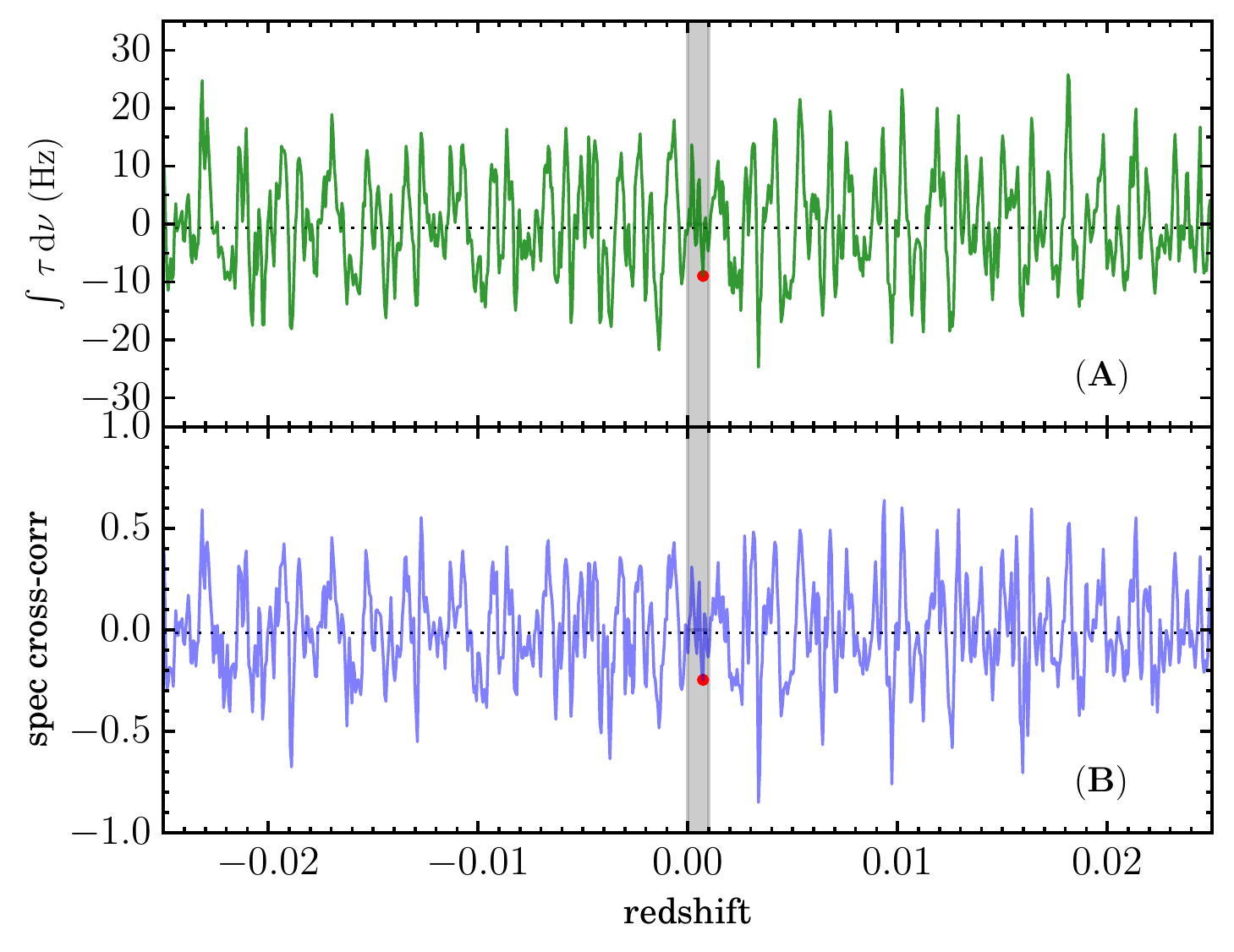}
    \caption{Same as in Figure \ref{fig:m82_rp} where we have implemented a stacking and spectral cross-correlation across a range of redshifts, except in this case the spectra have been combined and smoothed via a Savitzky-Golay filter. We mimicked the procedure \citetalias{Morabito2014} used to flag, process, and smooth the data of M~82 as best as possible.}
    \label{fig:m82_rl_zrange}
\end{figure}

As a final test, we assume the same flagging and processing criteria of \citetalias{Morabito2014} and apply our three step method. We make an RRL stack at each redshift (same redshift range described in Section \ref{sec:m82}). Instead of integrating within a region half the size of the line-blanking region ($\pm 25$ \kms), we taylor this to specifically match the FWHM of the line (as it will maximize the signal-to-noise of the feature) and integrate within $\pm 18$ \kms. The results across redshift are shown in Figure \ref{fig:m82_rl_zrange}. The mean is -0.1 Hz and the standard deviation of the values across redshift is 8.7 Hz. This agrees with the 8.8 Hz expectation from noise of $1.3 \times 10^{-3}$ integrated over 36 \kms\ at 56.5 MHz. The integrated optical depth at $z=0.00073$, which is 9.0 Hz, is also consistent with noise. As the second step of our method, we apply the spectral cross correlation. We find the cross-correlation value to be 1.0$\sigma$ in comparison to the distribution. We note that relative to $z=0.00070$, we find an integrated optical depth of 11.2 Hz and a cross-correlation value that is 1.2 times the standard deviation of the distribution. 

In conclusion, we find that smoothing subband spectra with a Savitzky-Golay filter and implementing the flagging procedure of \citetalias{Morabito2014} produces a stacked CRRL feature centered at $z=0.00070$ with a significance of 2.2$\sigma$. Comparing the integrated optical depth at this redshift with the value obtained across a range of redshifts produces a value that is 1.0$\sigma$.

\end{appendix}

\end{document}